\documentclass[12pt]{iopart}
\usepackage{amsfonts} 
\usepackage{graphicx}
\usepackage{xcolor}
\usepackage[export]{adjustbox}
\usepackage[labelformat=simple]{subcaption} 
\usepackage[square, numbers, sort&compress]{natbib}

\eqnobysec

\begin{document}
\title[Directed graphs and interferometry]{Directed graphs and interferometry
}
\author{Bruno Melo$^1$, Igor Brand\~ao$^1$, Carlos Tomei$^2$ and
Thiago Guerreiro$^1$
}

\address{$^1$Departamento de F\'isica, Pontif\'icia Universidade Cat\'olica do Rio de Janeiro}
\address{$^2$Departamento de Matem\'atica, Pontif\'icia Universidade Cat\'olica do Rio de Janeiro}

\eads{\mailto{brunomelo@aluno.puc-rio.br}, \mailto{igorbrandao@aluno.puc-rio.br},  \mailto{tomei@mat.puc-rio.br}, \mailto{barbosa@puc-rio.br}}

\vspace{10pt}
\begin{indented}
\item[]February 2020
\end{indented}

\begin{abstract}
The observed output of an interferometer is the result of interference among the parts of the input light beam traveling along each possible optical path. In complex systems, writing down all these possible optical paths and computing their cumulative effect can become a difficult task. We present an intuitive graph-based method for solving this problem and calculating electric fields within an interferometric setup, classical and quantum. We show how to associate a weighted directed graph to an interferometer and define rules to simplify these associated graphs. Successive application of the rules results in a final graph containing information on the desired field amplitudes. The method is applied to a number of  examples in cavity optomechanics and cavity-enhanced interferometers.

\end{abstract}


\maketitle

\tableofcontents

\newpage

\section{Introduction} 

Diagrammatic methods are ubiquitous in physics. Graphs play an important role as a calculational device in various fields of physics such as electrodynamics \cite{Curtis1998}, chemistry \cite{Gutman1986, Rouvray1996}, quantum field theory \cite{Feynman1965, Helvang2013}, topological matter \cite{kauffman1991}, statistical mechanics \cite{Kardar2007} and its application to network theory \cite{Albert2002}, as well as quantum information science \cite{Nielsen2006, Rossi2013}. The combinatorial aspects of graphs make them suitable for studying problems involving scattering \cite{Gibson2018, Gibson2019}, quantum optics \cite{Krenn_2017, Gu_2019} and quantum interferometry \cite{Ataman2015ThreeExperiments, Ataman2015Fabry, Ataman_2018}. In the present work we explore a graphical-based method, closely related to the one proposed by \cite{Ataman2014First}, aimed at analysing complex optical interferometers.

Interferometers can be viewed as a controlled scattering experiment. Calculating the transmission and reflection coefficients as well as the electric field at arbitrary positions of interferometric devices may not be a straightforward task, usually involving long matrix computations which obscure physical intuition. 
We explore an intuitive graph-based method for calculating such fields by drawing and manipulating pictures according to pre-determined rules. To any given homodyne linear optical setup, a weighted directed graph may be associated. All possible optical paths leading to a desired position in the interferometer are represented as walks in the graph and must be taken into account, much in the spirit of the Feynman integral. This graph can be simplified by successive application of the rules, and the resulting weighted directed graph contains the information on any transmission and reflection coefficients one wishes to obtain for the given interferometer. Similar ideas can be used both for classical and quantum fields.

The proposed rules make the graphical-based method suitable for a number of different applications. In this work we will focus on optical cavities, cavity-enhanced interferometers \cite{paper_LIGO} and optomechanical setups \cite{Aspelmeyer2014}, consisting of cavities with multiple dispersive elements \cite{Newsom2019, Bhattacharya2008}. It can also be applied to layered optical media such as photonic crystals \cite{Saleh2007}, quantum communication networks \cite{Osorio2012, Sangouard2011, Tillman2015} and the interferometric preparation of complex quantum states \cite{Krenn2016}.

This work is divided as follows. In section 2, it is shown, with the aid of examples, how to construct a weighted directed graph from a linear optical setup. It is then defined, in section 3, the general simplification rules to transform a graph. To demonstrate the power of the defined rules, a number of examples on how to calculate the transmitted field in complex optical setups using graph simplification is provided in section 4. Then, in section 5, the application of the method to calculate an intermediate field at an arbitrary point of a setup is discussed. The extension of the method to setups containing multiple inputs and outputs is presented in section 6. Section 7 deals with the application of the method to arbitrary quantum states of the electromagnetic field. The work is concluded in section 8 with final considerations.

\section{From an optics schematic to a weighted directed graph}

\begin{figure}[t]
\centering
\subcaptionbox{\ \label{fig:Michelson_Schematic}}
{\includegraphics[scale=0.8]{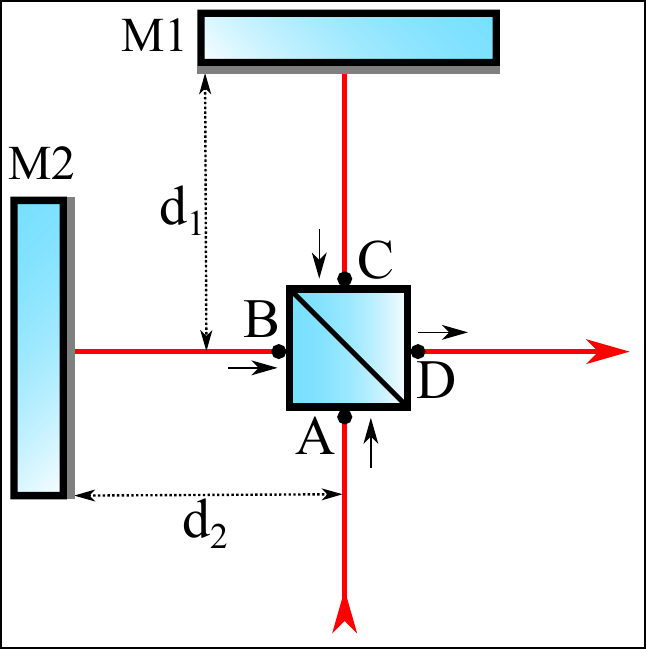}}
\subcaptionbox{\ \label{fig:Michelson_Diagram}}
{\includegraphics[scale = 0.8]{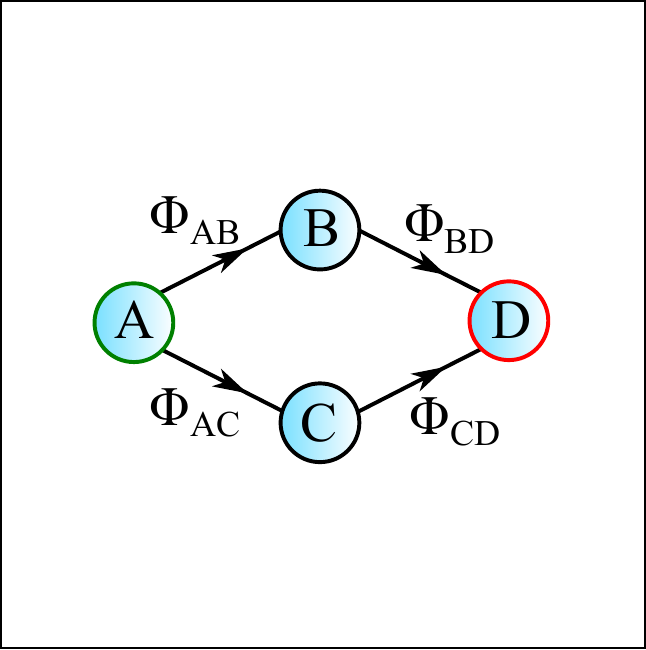}}
\caption{(a) Optical schematics of a Michelson interferometer. Black dots and arrows define the position and direction, respectively, of states A, B, C and D. The red lines represent light passing by the interferometer. (b) Directed graph G corresponding to the Michelson interferometers. The green and the red circles indicate, respectively, the input and the output vertices.}
\end{figure}
We describe, with the aid of examples, how to obtain a weighted directed graph from a standard optical schematics. Consider the Michelson interferometer in figure \ref{fig:Michelson_Schematic}. 
Each arrow, labeled by a capital letter, represents a \textit{state}, defined by a position and a direction in space. For instance, state $ \rm{B} $ in figure \ref{fig:Michelson_Schematic} represents a plane wave-front at position $ \rm{B} $ moving to the right. 

Each state is represented by a vertex in a directed graph, as seen in figure \ref{fig:Michelson_Diagram}.
States should be defined such that: (i) every optical path leading from the input to the output can be represented by a sequence of states and (ii) different optical paths are represented by different sequences of states.

An edge $\alpha_{ij}$ joins vertex $i$ to vertex $j$ if, and only if, the wave-front can go from state $i$ to state $j$ without passing by any other state along the way. To each edge $\alpha_{ij}$ a weight $\Phi_{ij}$ - the transition amplitude from state $i$ to state $j$ - is assigned. 

In figure \ref{fig:Michelson_Schematic}, for a wave-front with wave number $k$ in state A, two things can happen. First, it might get transmitted by the beam splitter (BS), with transmittance $t$ (and reflectance $r$), and then reflected back by the perfect mirror M1, ending up in state C. The transition amplitude for this process is the weight $\Phi_{\rm{AC}}=ite^{2ikd_1}$, and an edge connects the vertices A and C in the graph of figure \ref{fig:Michelson_Diagram}. Second, the wave-front might be reflected by the BS and the perfect mirror M2, resulting in state B; the edge connecting A and B has weight $\Phi_{\rm{AB}}=re^{2ikd_2}$.

Light in state C might also be reflected by the BS, resulting in the state D with transition amplitude $\Phi_{\rm{CD}}=r$. Similarly, light in state $ B $ can be transmitted to D, with the associated amplitude $\Phi_{\rm{BD}} = it $.

Since only the amplitude of the field in state D is of interest, there is no need to define a counter propagating state at A.  The resulting graph is shown in figure \ref{fig:Michelson_Diagram}, with all vertices, edges and weights depicted.

Each optical path from the input to the output of the interferometer corresponds in figure \ref{fig:Michelson_Diagram} to a walk from the source vertex, marked with a green outline, to the sink vertex, marked with a red outline. The \textit{weight} of the walk, which is the product $\Gamma_{\alpha}$ of the weights of all the edges along it, relates input and output electric fields  $\vec{E}_{in}$ and $\vec{E}_{out,\alpha}$ after the wave has traveled through the path

\begin{equation}
    \vec{E}_{\rm{out},\alpha} = \Gamma_\alpha \ \vec{E}_{\rm{in}}.
\end{equation}

The resultant electric field at the output is then given by

\begin{equation}
    \vec{E}_{\rm{out}} = \Gamma \ \vec{E}_{\rm{in}},
\end{equation}
\noindent where the response factor $\Gamma$ is the sum of the weights $\Gamma_\alpha$ of all walks from A to D.

In the example of figure \ref{fig:Michelson_Diagram}, there are only two possible walk: $\alpha_{\rm{AB}},\alpha_{\rm{BD}}$ and $\alpha_{\rm{AC}},\alpha_{\rm{CD}}$, of respective weights $\Phi_{\rm{AB}}\Phi_{\rm{BD}}$ and $\Phi_{\rm{AC}}\Phi_{\rm{CD}}$. The response factor is
\begin{equation}
    \Gamma = \Phi_{\rm{AB}}\Phi_{\rm{BD}} + \Phi_{\rm{AC}}\Phi_{\rm{CD}},
\end{equation}
\noindent yielding the known result for the electric field at the output of a Michelson interferometer
\begin{equation}
    \vec{E}_{\rm{out}} = irt(e^{2ikd_1}+e^{2ikd_2})\,\vec{E}_{\rm{in}}.
\end{equation}

\begin{figure}[t]
\centering
\subcaptionbox{\ \label{fig:Fabry_Schematics}}
{\includegraphics[scale = 0.8]{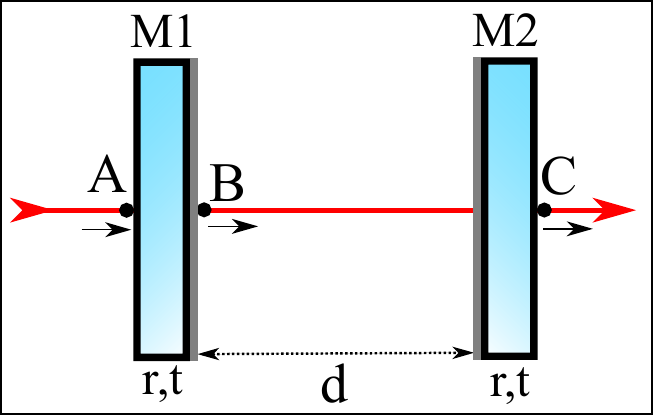}}
\subcaptionbox{\ \label{fig:Fabry_Diagram}}
{\includegraphics[scale = 0.8]{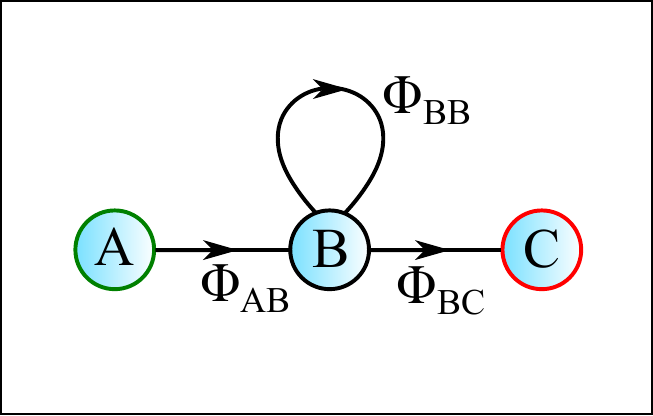}}
\caption{(a) Optical schematics of a Fabry-P\'erot cavity. Since in this example the cavity's reflection is not of interest, only the output state C is defined. (b) Graph G corresponding to the Fabry-P\'erot cavity. The loop in vertex B corresponds to the situation in which the wave-front undergoes one round-trip inside the cavity, going from state B back to state B.}
\end{figure}

It's important to point out that throughout this work the  convention in \cite{Jayich2008, paper_2_membranes} for the phase gained by reflected and transmitted waves is used. Moreover, only the case of monochromatic electric field with wave number $k$ is considered. 

As a second example, consider the Fabry–P\'erot interferometer in figure \ref{fig:Fabry_Schematics} with  states A, B and C. The equivalent graph is shown in figure \ref{fig:Fabry_Diagram}, in which $\Phi_{\rm{AB}}=it$, $\Phi_{\rm{BB}}=r^2e^{2ikd}$, $\Phi_{\rm{BC}}=ite^{ikd}$. The edge $\alpha_{BB}$ corresponds to a round trip inside the cavity.

Every walk from $A$ to $C$ starts with the edge $\alpha_{AB}$, continues with a number $n$, $n=0,1,\ldots \infty$, of loops $\alpha_{BB}$ and ends with the edge $\alpha_{BC}$. Adding up, since $|\Phi_{\rm{BB}}|<1$, 

\begin{equation}
\label{eq:faby_phi}
    \Gamma = \Phi_{\rm{AB}} \left(\sum_{n=0}^{\infty} \Phi_{\rm{BB}}^n \right) \Phi_{\rm{BC}} = \frac{\Phi_{\rm{AB}}\Phi_{\rm{BC}}}{1-\Phi_{\rm{BB}}}.
\end{equation}

Substituting the values for $\Phi_{\rm{AB}}$, $\Phi_{\rm{BB}}$ and $\Phi_{\rm{BC}}$ in (\ref{eq:faby_phi}) yields the usual expression $ \vec{E}_{\rm{out}}$ for the transmission of a Fabry–P\'erot cavity \cite{Ismail2016},
\begin{equation}
 \vec{E}_{\rm{out}} =  -\frac{t^{2} e^{ikd} }{1- r^2e^{2ikd}}\,\vec{E}_{\rm{in}}.
\end{equation}

\section{General rules for graph simplification}\label{sec:rules}
Before analyzing other interferometers, some {\it local simplification rules} are presented. As in electrical circuits, elements in series and/or in parallel are amenable to equivalent substitutions. This is the content of the first two rules. We say that graphs $\rm{G}$ and $\rm{\hat{G}}$ are {\it equivalent} if they have the same factor $\Gamma$ relating input and output, $ \vec{E}_{\rm{out}} =   \Gamma \vec{E}_{\rm{in}}$.

\begin{figure}[t]
\centering
\subcaptionbox{\ \label{fig:series_states}}
{\includegraphics[scale = 0.8]{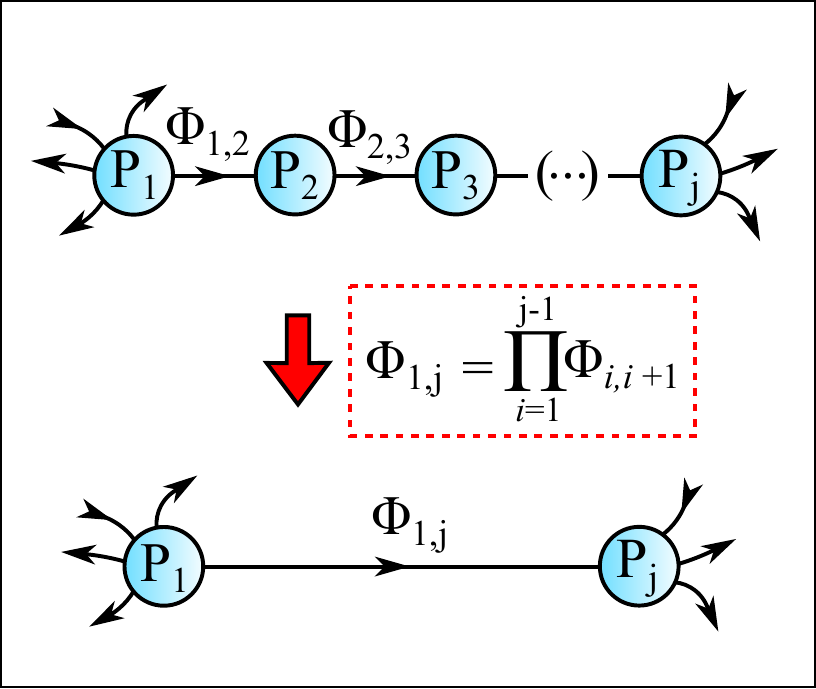}}
\subcaptionbox{\ \label{fig:parallel_states}}
{\includegraphics[scale = 0.8]{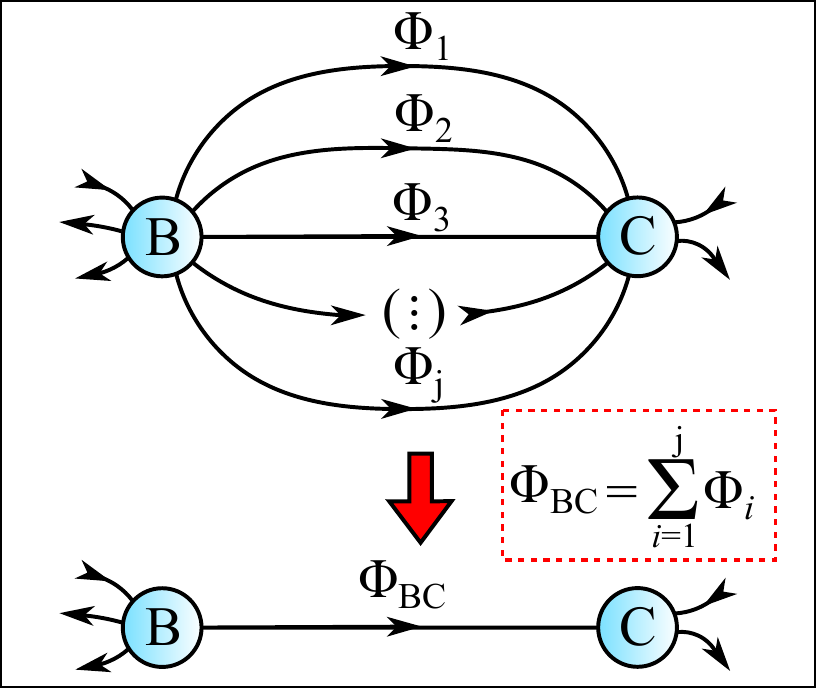}}
\caption{Graphs G and ${\rm{\hat{G}}}$ for two different rules in which multiple edges are replaced by a single edge with equivalent weight equal to (a) the product of the weights of each individual edge, if they are consecutive edges, or (b) the sum of the weights of each individual edge, if they are in parallel.}\label{fig:furst_rules}
\end{figure}

\subsection{Consecutive edges}
Suppose a vertex $P_2$ is connected only to two other vertices: vertex $P_1$, by an incoming edge $\alpha_{1,2}$, and vertex $P_3$, by an outgoing edge $\alpha_{2,3}$. A walk from $P_1$ to $P_3$ passing by $P_2$ must contain $\alpha_{1,2}$ and $\alpha_{2,3}$, which contribute with a factor $\Phi_{1,2}\Phi_{2,3}$ to the weight of the walk. An equivalent graph $\rm{\hat{G}}$  is obtained by removing the vertex $P_2$  and joining $P_1$ and $P_3$ by an edge $\alpha_{1,3}$ of weight $\Phi_{1,2}\Phi_{2,3}$. Similarly, for an arbitrary number of consecutive edges as shown in figure \ref{fig:series_states},  consecutive edges may be replaced by a single edge of weight given by the product of the weights of the individual edges.

\subsection{Parallel edges} \label{parallel}
Consider now a graph G with two vertices B and C joined by j different edges $\alpha_i$, with common orientation, and respective weights $\Phi_i$, as in figure  \ref{fig:parallel_states}.
Take $\rm{\hat{G}}$ to be the graph obtained from G replacing these edges by a single one, $\alpha_{\rm{BC}}$, of weight $\sum_i \Phi_i$. We show that G and $\rm{\hat{G}}$ are equivalent.

Each walk in G gives rise to a monomial given by the product of its edge weights. Each walk in $\rm{\hat{G}}$, instead, gives rise to a number of such monomials. It turns out that there is a simple bijection between equal monomials related to both graphs. Indeed, suppose that in G the walk goes from B to C $k$ times, with a contribution $\Phi_{i_1}\Phi_{i_1} \ldots  \Phi_{i_k}$ to the overall weight of the walk. Such walk in G corresponds to a walk in $\rm{\hat{G}}$ where  edge $\alpha_{BC}$ is traversed $k$ times, contributing with $(\sum_i \Phi_i)^k$ to the weight of the walk. The monomial  $\Phi_{i_1}\Phi_{i_2} \ldots  \Phi_{i_k}$ is naturally associated with the monomial in $(\sum_i \Phi_i)^k$ obtained by collecting $\Phi_{i_1}$ in the first term $\sum_i \Phi_i$, $\Phi_{i_2}$ in the second term, $\ldots$ , $\Phi_{i_k}$ in the $k$-th term.

The argument does not require that all edges between B and C have the same orientation. Collect edges with different orientations in two sets, and each set is replaced by a single edge as defined above.

\begin{figure}[t!]
\centering
\subcaptionbox{\ \label{fig:loop_states}}
{\includegraphics[scale = 0.8]{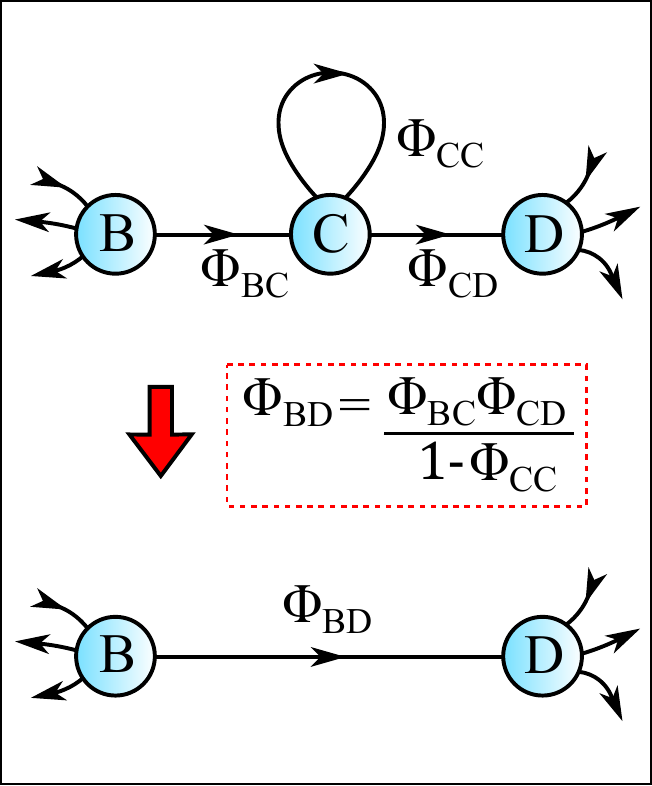}}
\subcaptionbox{\ \label{fig:many_loops_states}}
{\includegraphics[scale = 0.8]{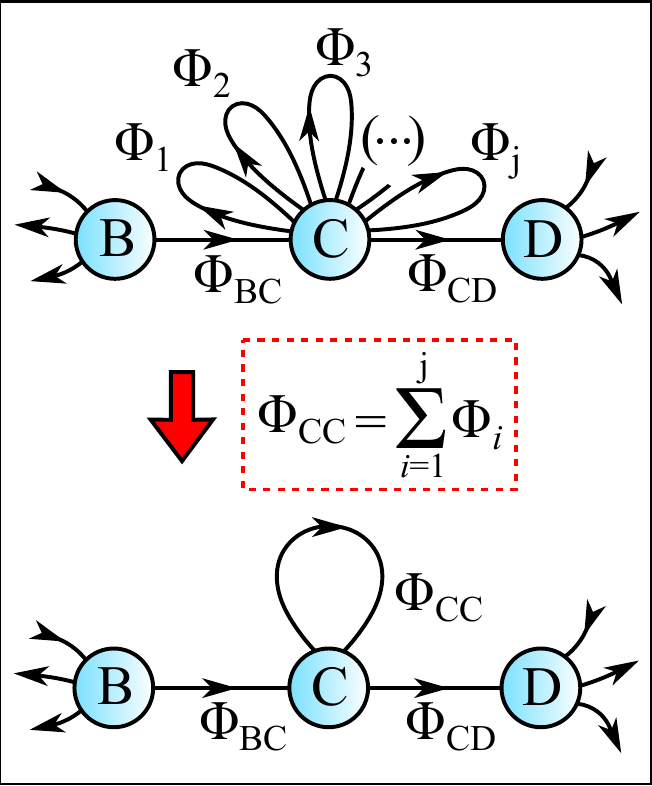}}
\caption{(a) A loop in vertex C, connected only to vertices B and D, is contracted by joining B and D with an equivalent edge. (b) Multiple loops at a vertex C can be replaced by a single loop with weight equal to the sum of the weights of each individual loop.}\label{fig:loop_rules}
\end{figure}

\subsection{Loop contraction}
Let a vertex C, that contains a loop, be connected to only two other vertices B and D, as in figure \ref{fig:loop_states}. The simplification rule in this case follows the argument for the Fabry-P\'erot interferometer above: the vertex C and its adjacent edges are eliminated and a single edge $\alpha_{\rm{BD}}$ is left with weight  $\Phi_{\rm{BC}}\Phi_{\rm{CD}}/(1-\Phi_{\rm{CC}})$, provided that $\vert \Phi_{\rm{CC}}\vert<1$.

Take now a graph with $\rm{j}$ loops at C with weights $\Phi_1, \ldots, \Phi_{\rm{j}}$, as in figure \ref{fig:many_loops_states}. As in the simplification of parallel edges, an equivalent graph $\rm{\hat {G}}$ is obtained by removing all but one loop, for which we assign weight $\sum_i \Phi_i$.

\begin{figure}[t!]
    \centering
    \includegraphics[scale = 0.8]{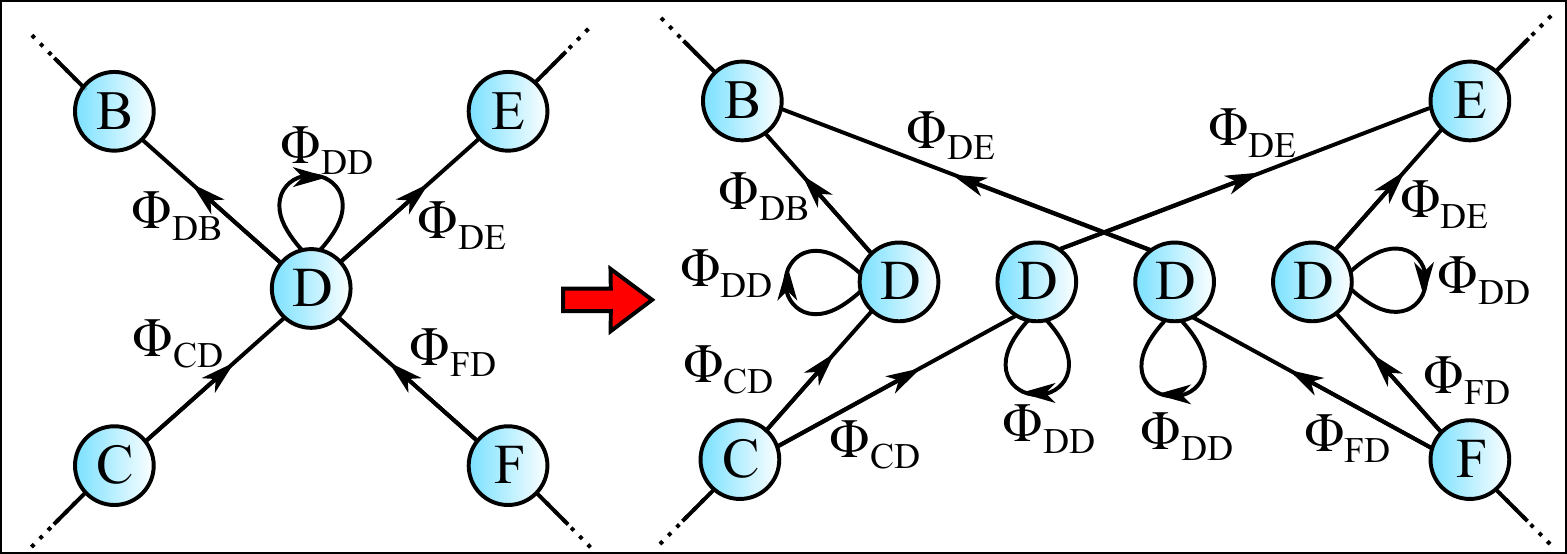}
    \caption{Vertex detaching: $\rm{\hat G}$ is obtained by creating copies of the vertex D so that walks are conserved; each copy of D has a single incoming and outgoing edge, which allows for the application of previous rules.}
    \label{fig:multiple_node}
\end{figure}

\subsection{Vertex detaching}
As a final rule, consider a graph G where a vertex D has $i$ incoming edges, $o$  outgoing edges and $\ell$ loops attached to it. From the simplification rule for loops, $\ell = 1$ without loss of generality; also, $i, o \ne 0$. In figure \ref{fig:multiple_node}, $i=2$, $o=2$, $\ell = 1$. The equivalent graph $\rm{\hat{G}}$ is obtained by replacing D by $i \cdot o$ vertices ${\rm{D}}_{m,n}, m = 1, \ldots i, n= 1, \ldots o$ with single incoming and outgoing edges (and the $\ell$ loops) such that the $i \cdot o$ pairs (incoming edge, outgoing edge) through D are reproduced in the $i \cdot o$ copies ${\rm{D}}_{m,n}$. 

To show the equivalence of G and $\rm{\hat{G}}$,  we again present a bijection between the sets of walks $\{ w \}$ in G and $\{ \hat w \}$ in $\rm{\hat{G}}$ which preserves the weight of each walk. Given $w$ in G, the corresponding  $\hat w$ is constructed by performing an alteration in $w$ whenever it passes through D. Each pass belongs to a short stretch ${\rm{I}}_m {\rm{D}} {\rm{O}}_n$ for unique vertices ${\rm{I}}_m$, from which the $m^{th}$ edge comes, and ${\rm{O}}_n$, to which the $n^{th}$ edge goes. The pass contributes to the weight of $w$ with $\Phi_{{\rm{I}}_m {\rm{D}}} \Phi_{{\rm{D}} {\rm{O}}_n}$. To obtain $\hat w$, replace the stretch ${\rm{I}}_m {\rm{D}} {\rm{O}}_n$ by a stretch ${\rm{I}}_m {\rm{D}}_{m,n} {\rm{O}}_n$ joining the vertices of $\rm{\hat{G}}$ and preserve loops, if any. Clearly, the construction yields the desired bijection.

As a final remark, notice that all simplification rules are {\it local}: they are performed in a very limited region of the graph, and absolutely do not depend on the graph outside of this region. Thus for example, vertex detaching is circumscribed to one vertex (D, in the example above) and the edges which contain it. The reader will have no difficulty in identifying the appropriate region of each simplification rule.

\section{Application to optical cavities}

Optical cavities are frequently used to increase the circulating power in a interferometer \cite{Webb1998a, Sato2000} and enhance sensitivity in displacement measurements \cite{Wei2019, Magrini2018, Vitali2002}. Since cavities give rise to an infinite number of possible optical paths, summing the amplitude of the waves undergoing each possible path might become a challenging task. To show how the graph-based method can handle this type of calculation, different systems containing optical cavities are studied with it.

\begin{figure}[t]
\centering
\subcaptionbox{\ \label{fig:2_membranes_schematic}}
{\includegraphics[scale = 0.7]{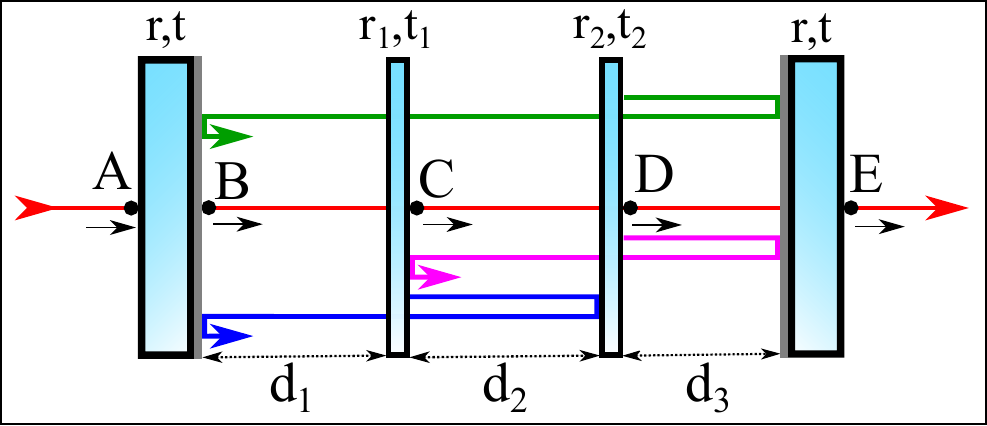}}
\subcaptionbox{\ \label{fig:2_membranes_diagram}}
{\includegraphics[scale = 0.7]{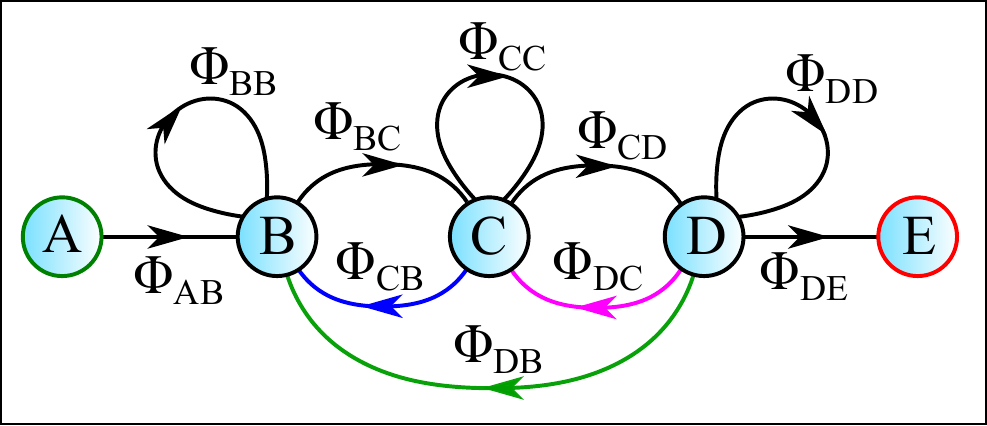}}
\caption{(a) Optical schematics of a cavity with two membranes in the middle. The red line represents light passing by the cavity, while the blue, green and pink lines indicates different optical paths inside the cavity. (b) Graph corresponding to a cavity with two membranes in the middle. The blue, green and pink edges correspond to the blue, green and pink optical paths displayed in figure \ref{fig:2_membranes_schematic}, respectively.}\label{fig:2_membranes}
\end{figure}

\subsection{Two membranes inside an optical cavity}
In cavity optomechanics, one interesting setup is a Fabry-P\'erot cavity with thin membranes positioned inside of it. The membranes, typically made of Silicon Nitride (Si$_{3}\rm{N}_{4}$), act as dispersive optical elements, and change the cavity resonance frequency according to where they are positioned with respect to the cavity's nodes \cite{Jayich2008, Thompson2008}.

The case of two membranes in the optical cavity is represented in figure \ref{fig:2_membranes_schematic}. The first [second] membrane's reflectance and transmittance are $r_1,t_1$ [$r_2,t_2$], respectively, while the cavity's mirrors are identical and have reflectance $r$ and transmittance $t$.

State A is defined where the electric field enters the optical cavity, and states B, C, D and E are defined just after each optical element, all with the same direction as the incident electric field. The associated graph, constructed according to the prescription of the above sections, is shown in figure \ref{fig:2_membranes_diagram}. The weights present in the graph are given, in terms of the element's reflection and transmission coefficients, by 
 \begin{eqnarray}
     \Phi_{\rm{AB}} = it && \;\;\;\;\Phi_{\rm{BC}} = it_1\e^{ikd_1}\nonumber \\ \Phi_{\rm{BB}} = r_1 r \e^{ik2d_1} && \;\;\;\;
     \Phi_{\rm{CC}} = r_2r_1\e^{ik2d_2} \nonumber \\ \Phi_{\rm{CD}} = it_2\e^{ikd_2} && \;\;\;\;
     \Phi_{\rm{CB}} = r_2it_1r\e^{ik(2d_2+d_1)} \nonumber\\ \Phi_{\rm{DB}} = rit_2it_1r\e^{ik(2d_3+d_2+d_1)} &&  \;\;\;\;\Phi_{\rm{DC}} = rit_2r_1\e^{ik(2d_3+d_2)} \nonumber\\\label{eq:two_membrane_weights} \Phi_{\rm{DD}} = r r_2\e^{ik2d_3} && \;\;\;\; \Phi_{\rm{DE}} = it \e^{ikd_3} 
 \end{eqnarray}

Note that the optical path leading directly from state D to state B (green arrow in figure \ref{fig:2_membranes_schematic}) is different from the optical path leading from state D to C and then from C to B (pink arrow followed by blue arrow in figure \ref{fig:2_membranes_schematic}). 

\begin{figure}[t]
    \centering
    \includegraphics[width=0.9\textwidth]{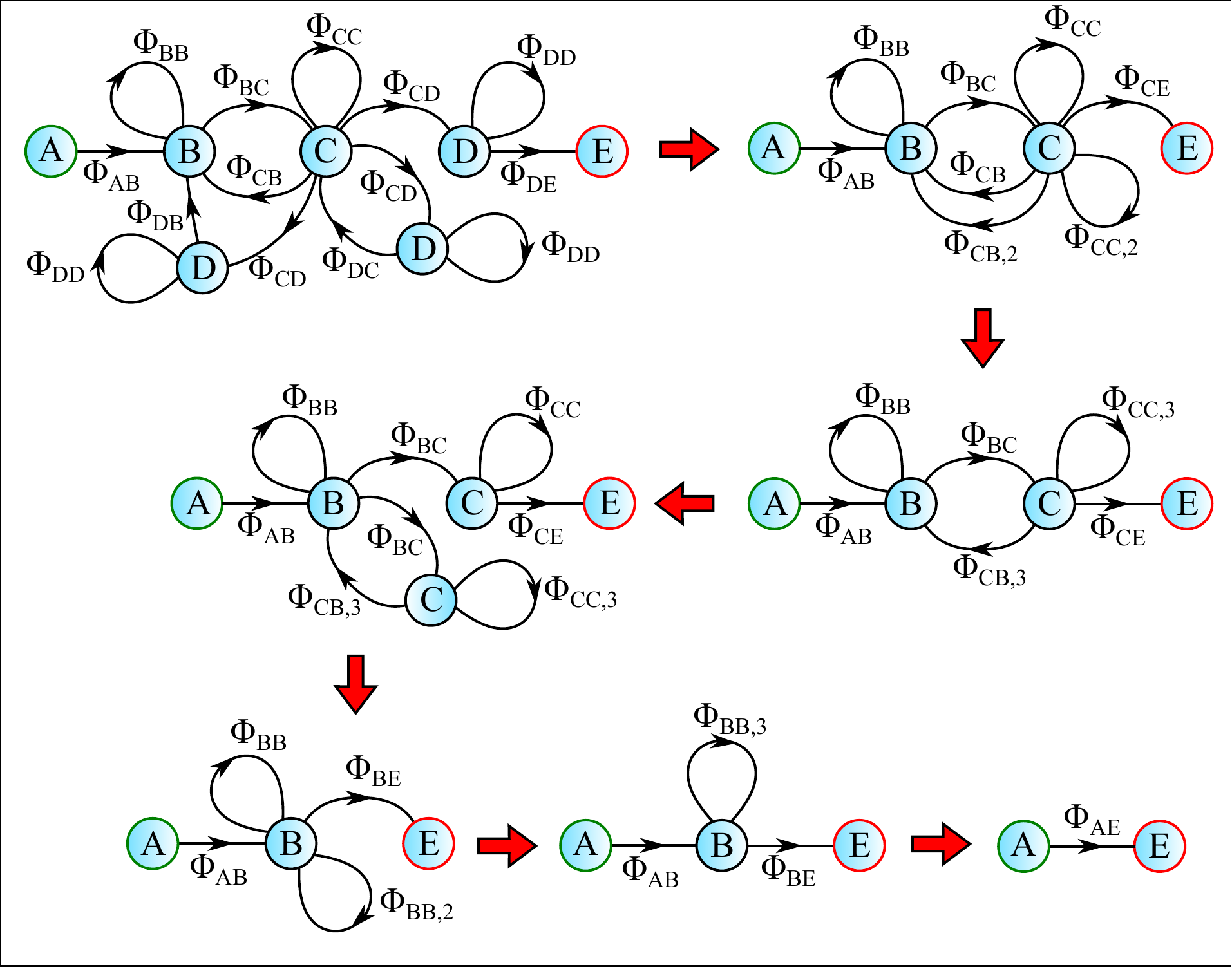}
    \caption{Successive application of simplification rules to the graph corresponding to the two membranes in the middle setup. The rules are applied until only the input and the output states are left. The weight of the edge connecting A to E in the final graph is equal to the response factor $\Gamma$.}
    \label{fig:2_membranes_simplification}
\end{figure}

In order to find the response factor of this setup, the simplification rules can be applied to the graph in figure \ref{fig:2_membranes_diagram}. 
First, perform vertex detaching on D, and find the first graph in figure \ref{fig:2_membranes_simplification}.
Then apply three loop contractions: one between vertices C and E, creating the edge $\alpha_{\rm{CE}}$ with weight $\Phi_{\rm{CE}} = \Phi_{\rm{CD}}\Phi_{\rm{DE}}/ (1-\Phi_{\rm{DD}})$; another between vertices C and C, creating the edge $\alpha_{\rm{CC},2}$ with weight $\Phi_{\rm{CC},2} = \Phi_{\rm{CD}}\Phi_{\rm{DC}}/ (1-\Phi_{\rm{DD}})$; and one more between vertices C and B, creating the edge $\alpha_{\rm{CB},2}$ with weight $\Phi_{\rm{CB},2} = \Phi_{\rm{CD}}\Phi_{\rm{DB}}/ (1-\Phi_{\rm{DD}})$. This leads to the second graph in figure \ref{fig:2_membranes_simplification}.
Now sum the loops in vertex C creating the loop $\alpha_{\rm{CC},3}$ with weight $\Phi_{\rm{CC},3} = \Phi_{\rm{CC}}+\Phi_{\rm{CC},2}$, and sum the parallel edges between vertices C and B, creating the edge $\alpha_{\rm{CB},3}$ with weight $\Phi_{\rm{CB},3} = \Phi_{\rm{CB}}+\Phi_{\rm{CB},2}$. This leads to the third graph in figure \ref{fig:2_membranes_simplification}; note that ignoring the weights, this graph is equivalent to the one associated to an optical cavity containing only one membrane. \par

Then, perform vertex detaching on C to get the fourth graph; eliminate the loop on each copy of the vertex C, as to create the edges $\alpha_{\rm{BE}}$ and $\alpha_{\rm{BB},2}$ with weights $\Phi_{\rm{BE}} = \Phi_{\rm{BC}}\Phi_{\rm{CE}}/ (1-\Phi_{\rm{CC},3})$ and $\Phi_{\rm{BB},2} = \Phi_{\rm{BC}}\Phi_{\rm{CB},3}/ (1-\Phi_{\rm{CC},3})$, respectively, and arrive at the fifth graph. After summing the loops on state B, creating the edge $\alpha_{\rm{BB},3}$ with weight $\Phi_{\rm{BB},3} = \Phi_{\rm{BB}} + \Phi_{\rm{BB},2}$, the sixth graph on figure \ref{fig:2_membranes_simplification} is obtained. Once again, ignoring the weights, this graph has the same structure as the graph for a cavity with no membranes in the middle.\par

Finally, eliminate the loop in B and find $\Phi_{\rm{AE}} = \Phi_{\rm{AB}}\Phi_{\rm{BE}}/ (1-\Phi_{\rm{BB},3})$, which is equal to the response factor between the input and output electric fields

\begin{eqnarray}
\label{eq:Two_Membranes_Phi}
 \Gamma = \frac{\Phi_{\rm{AB}}\,\Phi_{\rm{BC}}\,\Phi_{\rm{CD}}\,\Phi_{\rm{DE}}}
    {\Delta} \, ,
\end{eqnarray}
\noindent where
\begin{eqnarray}
\Delta \equiv  1-\Phi_{\rm{BB}}-\Phi_{\rm{CC}}-\Phi_{\rm{DD}}+
    \Phi_{\rm{BB}}\Phi_{\rm{CC}}+\Phi_{\rm{CC}}\Phi_{\rm{DD}}+\Phi_{\rm{DD}}\Phi_{\rm{BB}}\nonumber\\-
    \Phi_{\rm{CD}}\Phi_{\rm{DC}}-\Phi_{\rm{BC}}\Phi_{\rm{CB}}+
    \Phi_{\rm{BC}}\Phi_{\rm{CB}}\Phi_{\rm{DD}}+\Phi_{\rm{CD}}\Phi_{\rm{DC}}\Phi_{\rm{BB}}\nonumber\\-\Phi_{\rm{BC}}\Phi_{\rm{CD}}\Phi_{\rm{DB}}-\Phi_{\rm{BB}}\Phi_{\rm{CC}}\Phi_{\rm{DD}}\, .
\end{eqnarray}

\noindent Substituting the expressions in (\ref{eq:two_membrane_weights}) for the weights in terms of the reflectances and transmittances of each element in (\ref{eq:Two_Membranes_Phi}), the response factor becomes
\begin{eqnarray}
\label{eq:two_membranes}
\Gamma = \frac{t^2\,t_{1}\,t_{2}\,\e^{ik(d_1+d_2+d_3)}}{\Delta} \, ,
\end{eqnarray}
\noindent where
\begin{eqnarray}
\Delta = -r^2\e^{2ik(d_1+d_2+d_3)}\bigg(r_1^2r_2^2+r_1^2t_2^2+r_2^2t_1^2+t_1^2t_2^2\bigg)\nonumber\\+\bigg(\e^{2ik(d_1+d_2)}(r_1^2r_2+r_2t_1^2)+\e^{2ik(d_2+d_3)}r_1(r_2^2+t_2^2)-\e^{2ikd_1}r_1-\e^{ikd_3}r_{2}\bigg)\,r\nonumber\\+\e^{2ik(d_1+d_3)}r_1r_2r^2 - \e^{2ikd_2}r_1r_2+1 \,,
\end{eqnarray}
which is in agreement with the well-known result for the electric field transmitted by the cavity \cite{paper_2_membranes, Li2016}.

\subsection{N membranes inside an optical cavity}

\begin{figure}[t]
    \centering
    \includegraphics[scale = 0.8]{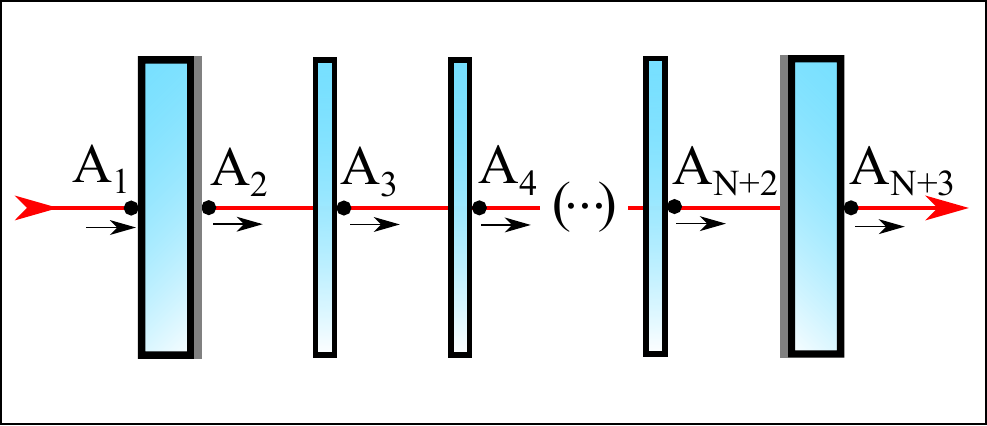}
    \caption{Schematics for $N$ membranes inside a cavity. Since one state is defined before the input mirror and one state is defined after each optical element, there is a total of $N+3$ states.}
    \label{fig:N_membranes_Schematics}
\end{figure}

As a generalization of the previous case, consider N membranes inside a cavity, as illustrated in figure \ref{fig:N_membranes_Schematics}. Whereas in previous works this system has been studied in terms of the optomechanical interaction provided by the membranes \cite{Bhattacharya2008, Newsom2019, Xu2013}, here the focus will be on deriving a method for calculating the system's transmission. To do so draw the graph formed by the $\rm{N}+3$ vertices corresponding to the states defined in figure \ref{fig:N_membranes_Schematics} and by the edges:
\begin{itemize}
    \item $\alpha_{k,k+1}$ for $1\leq k\leq \rm{N+2}$; 
    \item $\alpha_{k,j}$ for $2\leq j < k\leq  \rm{N+2}$;  
    \item $\alpha_{k,k}$ for $2\leq k \leq \rm{N+2}$; 
\end{itemize}
with $\alpha_{k,j}$ being the edge from ${\rm{A}}_{k}$ to ${\rm{A}}_{j}$. In particular, the state ${\rm{A}}_{\rm{N+2}}$ has
\begin{itemize}
    \item the incoming edge $\alpha_{\rm{N+1},\rm{N+2}}$;
    \item the outgoing edges $\alpha_{\rm{N+2}, k}$, for $2\leq k\leq \rm{N+1}$, and $\alpha_{\rm{N+2},\rm{N+3}}$;
    \item the loop $\alpha_{\rm{N+2},\rm{N+2}}$.
\end{itemize}

\begin{figure}[t]
\centering
\subcaptionbox{\ \label{fig:N_walk_1}}
{\includegraphics[scale = 0.8]{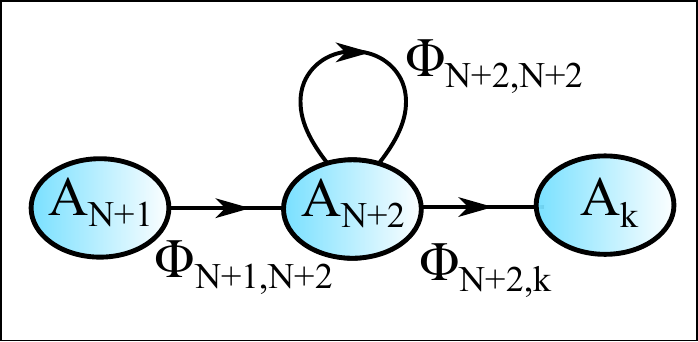}}
\subcaptionbox{\ \label{fig:N_walk_2}}
{\includegraphics[scale = 0.8]{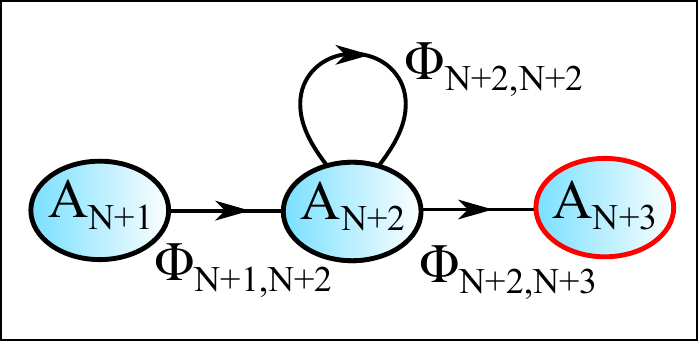}}
\caption{{Walks arising from the detachment of the state ${\rm{A}}_{\rm{N+2}}$}: (a) Walk from vertex $\rm{A}_{\rm{N+1}}$ to vertex $\rm{A}_k$, for $k$ such that $2\leq k\leq \rm{N+1}$. This walk and the edge $\alpha_{\rm{N+1},k}$ are in parallel. (b) Walk from vertex $\rm{A}_{\rm{N+1}}$ to the output vertex $\rm{A}_{\rm{N+3}}$.} \label{fig:N_walks}
\end{figure}

Detaching the state ${\rm{A}}_{\rm{N+2}}$ yields (i) one walk from ${\rm{A}}_{{{\rm{N}}+1}}$ to ${{\rm{A}}}_{k}$ as the one shown in figure \ref{fig:N_walk_1} for all $k$ such that $2\leq k\leq \rm{N+1}$ and (ii) one walk from ${\rm{A}}_{\rm{N+1}}$ to ${\rm{A}}_{\rm{N+3}}$ as the one shown in figure \ref{fig:N_walk_2}. Eliminating the loop in the walk from ${\rm{A}}_{\rm{N+1}}$ to ${\rm{A}}_k$, for $2\leq k\leq \rm{N+1}$, gives a new edge from ${\rm{A}}_{\rm{N+1}}$ to ${\rm{A}}_k$ that is in parallel with the initially existing edge $\alpha_{\rm{N+1},k}$. Merging these two edges yields the final weight for the edge from ${\rm{A}}_{\rm{N+1}}$ to ${\rm{A}}_k$:

\begin{equation}
    \Phi_{\rm{N+1},k}^{(\rm{N})}{'} = \Phi_{\rm{N+1},k}^{(\rm{N})}+\frac{\Phi_{\rm{N+1},\rm{N+2}}^{(\rm{N})}\Phi_{\rm{N+2},k}^{(\rm{N})}}{1-\Phi_{\rm{N+2},\rm{N+2}}^{(\rm{N})}}
\end{equation}
where the superscript $(\rm{N})$ emphasizes that the $\Phi$'s in this equation are the ones defined for the \textit{N membranes in the middle} case, whereas the prime symbol is used to distinguish the weight after the simplification from the weights before any change is made to the graph.

For the walk from ${\rm{A}}_{\rm{N+1}}$ to ${\rm{A}}_{\rm{N+3}}$, eliminating the loop yields an edge with weight
\begin{equation}
    \Phi_{\rm{N+1},\rm{N+3}}^{(\rm{N})}{'} =
    \frac{\Phi_{\rm{N+1},\rm{N+2}}^{(\rm{N})}\Phi_{\rm{N+2},\rm{N+3}}^{(\rm{N})}}{1-\Phi_{\rm{N+2},\rm{N+2}}^{(\rm{N})}}
\end{equation}
  
Renaming the vertex ${\rm{A}}_{\rm{N+3}}$ to ${\rm{A}}_{\rm{N+2}}$, the resulting graph has the same structure, although not the same weights, of a graph for $\rm{N}+1$ membranes: states ${\rm{A}}_k$, with $k$ ranging from $1$ to $\rm{N+2}$, and the edges
\begin{itemize}
    \item $\alpha_{k,k+1}$ for $1\leq k\leq \rm{N+1}$;
    \item $\alpha_{k,j}$ for $2\leq j < k\leq  \rm{N+1}$;
    \item $\alpha_{k,k}$ for $2\leq k \leq \rm{N+1}$.
\end{itemize}

\noindent Therefore, if the response factor $\Gamma^{(\rm{N-1})}$ is known for N-1 membranes as a function of the weights $\Phi^{(\rm{N-1})}_{i,j}$, the response factor $\Gamma^{(\rm{N})}$ for N membranes can be easily obtained by making the following substitutions
\begin{eqnarray}
\label{eq:substitutions1}
    \Phi_{\rm{N+1},\rm{N+2}}^{(\rm{N-1})} 
    &&\to
    \frac{\Phi_{\rm{N+1},\rm{N+2}}^{(\rm{N})}\Phi_{\rm{N+2},\rm{N+3}}^{(\rm{N})}}{1-\Phi_{\rm{N+2},\rm{N+2}}^{(\rm{N})}};\nonumber
    \\
\label{eq:substitutions2}
    \Phi_{\rm{N+1},\rm{N+2}}^{(\rm{N-1})} 
    &&\to
    \Phi_{\rm{N+1},k}^{(\rm{N})}
    +
    \frac{\Phi_{\rm{N+1},\rm{N+2}}^{(\rm{N})}\Phi_{\rm{N+2},k}^{(\rm{N})}}{1-\Phi_{\rm{N+2},\rm{N+2}}^{(\rm{N})}},\textrm{ for }2\leq k\leq \rm{N+1}; \nonumber
    \\
\label{eq:substitutions3}
    \Phi_{i,j}^{(\rm{N-1})} &&\to \Phi_{i,j}^{(\rm{N})}, \textrm{for the remaining weights}.
\end{eqnarray}

\begin{figure}[t]
    \centering
    \includegraphics[scale = 0.8]{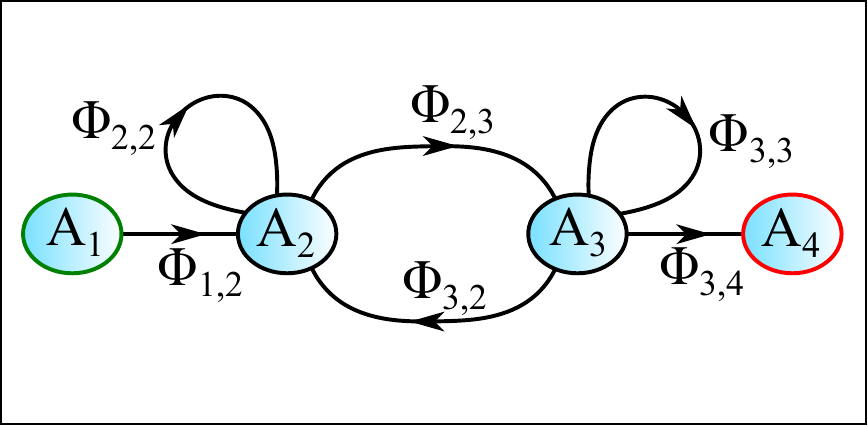}
    \caption{Graph associated to the transmission of a cavity with a membrane in the middle. Applying the simplification described in the present section to this graph yields the graph associated to a cavity with no membranes inside of it. Conversely, applying the substitutions prescribed in (\ref{eq:substitutions3}) to the response factor of a Fabry-Per\'ot cavity gives the response factor for the displayed graph.}
    \label{fig:Membrane_Diagram}
\end{figure}

The case of an empty Fabry-P\'erot cavity ($\rm{N-1}=0$) can be used as an example. The response factor for this cavity is given by
\begin{equation}
    \Gamma^{(0)} = \frac{\Phi_{1,2}^{(0)}\Phi_{2,3}^{(0)}}{1-\Phi_{22}^{(0)}}.
\end{equation}

Now, making the substitutions prescribed in (\ref{eq:substitutions3}), one arrives at

\begin{eqnarray}
 \Gamma^{(1)} &&= \frac{\Phi_{1,2} [ \Phi_{2,3} \Phi_{3,4}/(1-\Phi_{3,3}) ]}{1-[\Phi_{2,2}+\Phi_{2,3}\Phi_{3,2}/(1-\Phi_{3,3})]}\nonumber
 \\
 &&=\frac{\Phi_{1,2}\Phi_{2,3}\Phi_{3,4}}{1-\Phi_{2,2}-\Phi_{3,3}+\Phi_{2,2}\Phi_{3,3}-\Phi_{2,3}\Phi_{3,2}},
\end{eqnarray}
where the superscript $(1)$ in all the $\Phi$'s has been omitted. This is the right expression for the response factor of a cavity containing one membrane, as can be directly verified by calculating $\Gamma^{(1)}$ from the graph shown in figure \ref{fig:Membrane_Diagram}.

Note that in order to get the structure of the graph for N-1 membranes starting from the graph for N membranes,  N+1 loop eliminations followed by N merges of edges in parallel are necessary. Therefore, the number of operations in this procedure is of order $\mathcal{O}(N)$. By repeating this process it is possible to simplify the graph for N membranes to get a single edge connecting the initial and final vertices with a number of operations that is of order $\mathcal{O}(N^2)$.

\subsection{Cavity-enhanced Michelson interferometer}
\begin{figure}[t]
\centering
\subcaptionbox{Optical setup. \label{fig:LIGO_schematics}}
{\includegraphics[scale = 0.8]{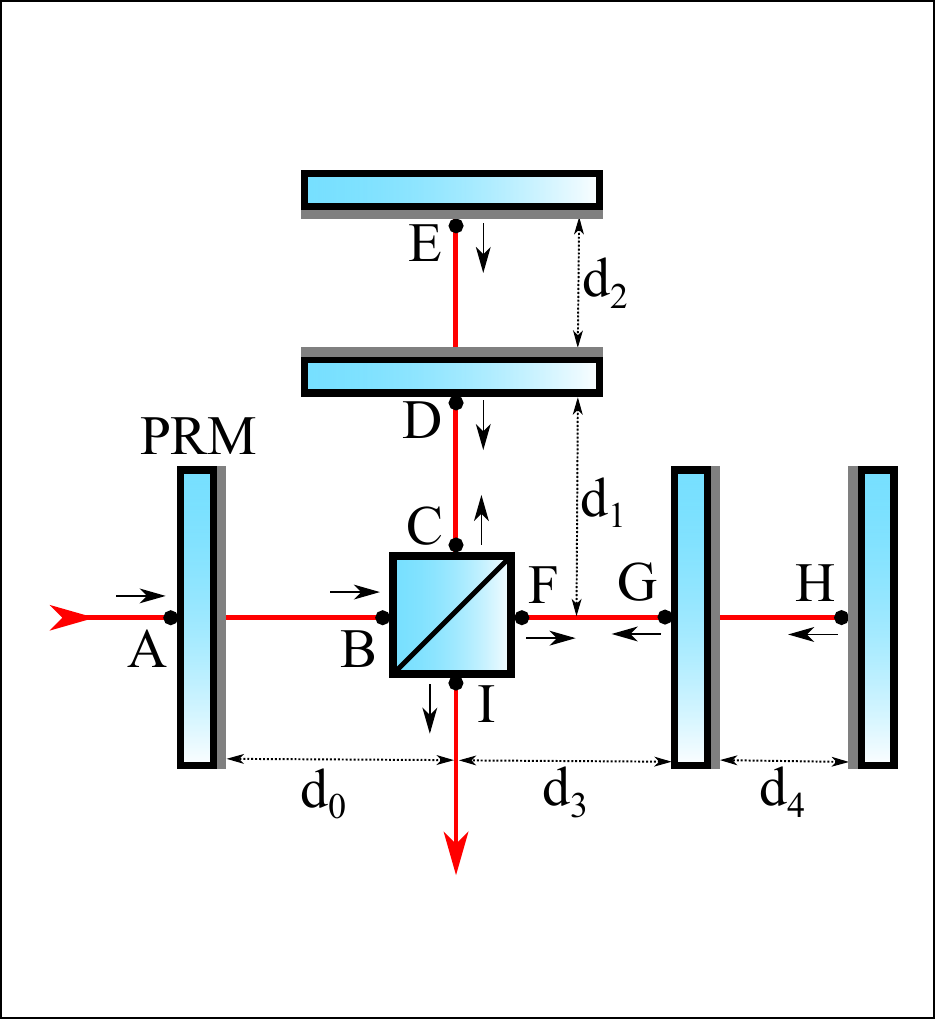}}
\subcaptionbox{The associated graph.\label{fig:LIGO_diagram}}
{\includegraphics[scale = 0.8]{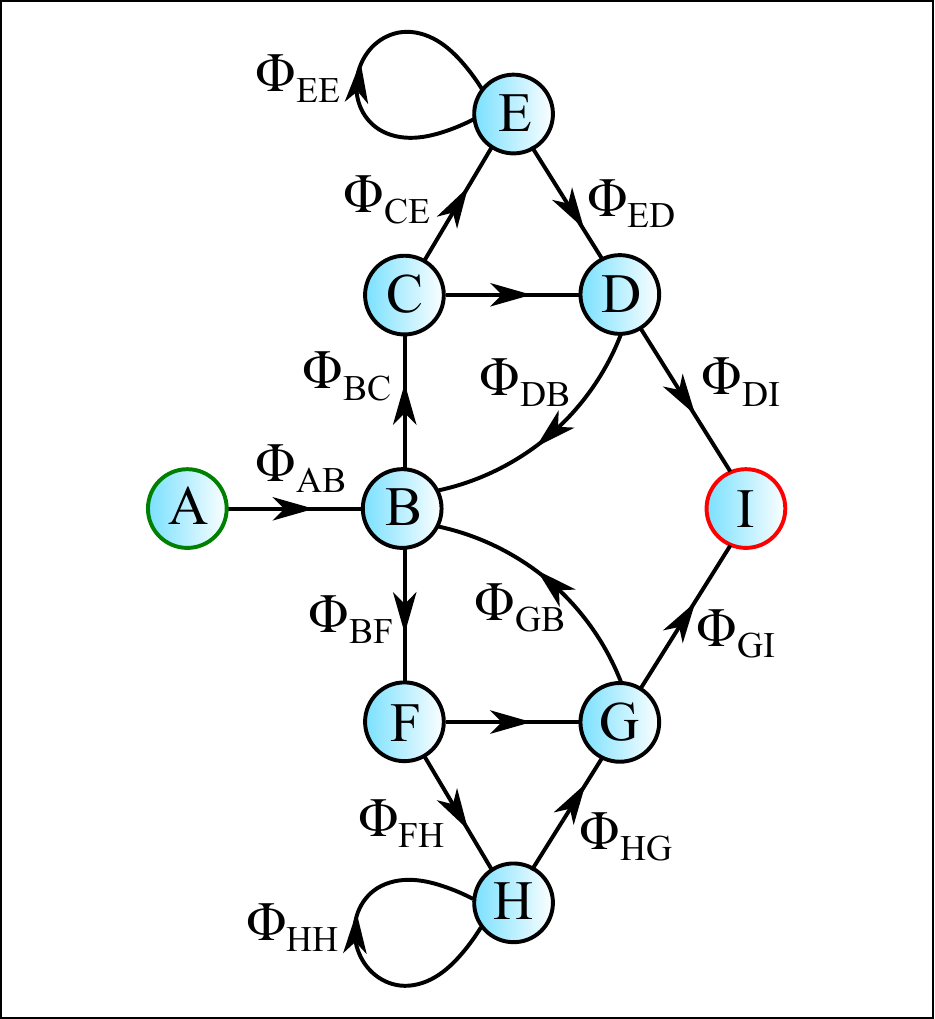}}
\caption{(a) Optical schematics of a cavity-enhanced Michelson interferometer: a cavity is placed at the end of each arm of a Michelson interferometer, in order to increase sensitivity, and a mirror is placed before the BS, in order to increase the power stored in the system. (b) Graph corresponding to the cavity-enhanced Michelson interferometer.}\label{fig:LIGO}
\end{figure}

One modern type of interferometer is the Michelson interferometer with cavities at the end of both arms and a power recycling mirror (PRM) at the input, as the one displayed in figure \ref{fig:LIGO_schematics}. This configuration, including similar ones \cite{Muler2003, Thuring2005}, can be used in gravitational-waves observatories such as the Laser Interferometer Gravitational Wave Observatory (the LIGO collaboration) \cite{Abramovici1992}.

The reflectance and transmittance are: $r_1,t_1$ for the PRM, $r_2, t_2$ for the BS and $r_3, t_3$ for the optical cavities' mirrors. Defined states A (input) through I (output), the corresponding graph is displayed in figure \ref{fig:LIGO_diagram} and the relevant transition amplitudes are given by

\begin{eqnarray}
\Phi_{\rm{AB}} = it_1\e^{ikd_0} && \;\; \Phi_{\rm{BC}} = r_2 \nonumber\\ 
\Phi_{\rm{BF}} = it_2 && \;\; \Phi_{\rm{CD}} = r_3\e^{ikd_1} \nonumber\\
\Phi_{\rm{CE}} = it_3r_3\e^{ik(d_1+d_2)} && \;\; \Phi_{\rm{EE}} = r_3^2\e^{ik2d_2} \nonumber\\
\Phi_{\rm{ED}} = it_3\e^{ikd_2}  && \;\; \Phi_{\rm{DI}} = it_2\e^{ikd_1} \nonumber\\
\Phi_{\rm{DB}} = r_1r_2\e^{ik(d_1 + 2d_0)}  && \;\; \Phi_{\rm{FG}} = r_3\e^{ikd_3} \nonumber\\
\Phi_{\rm{FH}} = it_3r_3\e^{ik(d_3+d_4)} && \;\; \Phi_{\rm{HH}} = r_3^2\e^{ik2d_4} \nonumber\\
\Phi_{\rm{HG}} = it_3\e^{ikd_4}  && \;\; \Phi_{\rm{GI}} = r_2\e^{ikd_3} \nonumber\\
\Phi_{\rm{GB}} = it_2r_1\e^{ik(d_3+2d_0)}
\end{eqnarray}

\begin{figure}[t]
    \centering
    \includegraphics[scale = 0.8]{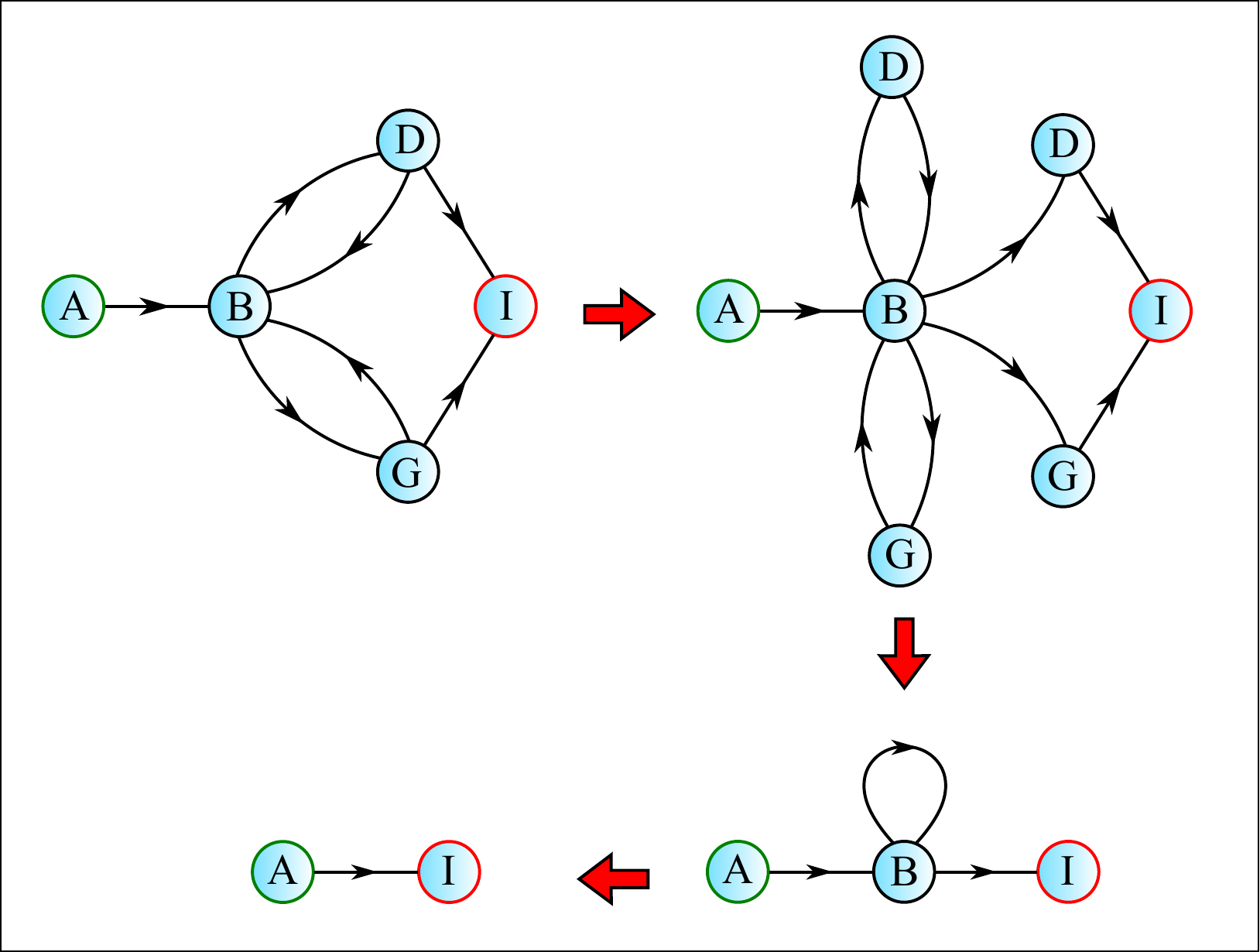}
    \caption{Successive application of simplification rules to the graph corresponding to a cavity-enhanced Michelson interferometer. As in the two membranes in the middle case, rules are applied until only the input and output vertices remain.}
    \label{fig:LIGO_simplification}
\end{figure}

To find the response factor of this interferometer, first eliminate the loops in H and E in figure \ref{fig:LIGO_diagram}, sum the resulting parallel edges, between F and G and between C and D, and multiply the consecutive edges between B and G and between B and D. This sequence of operations leads to the first graph in figure \ref{fig:LIGO_simplification}. Continue by detaching states D and G to arrive at the second graph in figure \ref{fig:LIGO_simplification}. Next multiply the consecutive edges $\alpha_{\rm{BD}},\alpha_{\rm{DB}}$ and $\alpha_{\rm{BG}},\alpha_{\rm{GB}}$, and sum the resulting loops in state B; multiply the consecutive edges between B and I and sum the resulting edges, arriving at the third graph in figure \ref{fig:LIGO_simplification}.

All that remains is a loop contraction at B for one to arrive at the final graph in figure \ref{fig:LIGO_simplification}, yielding the following output response factor

\begin{equation}
    \Gamma =  \frac{\Phi_{\rm{AB}}(\Phi_{\rm{BF}}\big( \frac{\Phi_{\rm{FH}}\Phi_{\rm{HG}}}{1-\Phi_{\rm{HH}}}+\Phi_{\rm{FG}} \big)\Phi_{\rm{GI}}+\Phi_{\rm{BC}}\big( \frac{\Phi_{\rm{CE}}\Phi_{\rm{ED}}}{1-\Phi_{\rm{EE}}}+\Phi_{\rm{CD}} \big)\Phi_{\rm{DI}})}{1 - (\Phi_{\rm{BF}}\big( \frac{\Phi_{\rm{FH}}\Phi_{\rm{HG}}}{1-\Phi_{\rm{HH}}}+\Phi_{\rm{FG}} \big)\Phi_{\rm{GB}} + \Phi_{\rm{BC}}\big( \frac{\Phi_{\rm{CE}}\Phi_{\rm{ED}}}{1-\Phi_{\rm{EE}}}+\Phi_{\rm{CD}} \big)\Phi_{\rm{DB}})} .
\end{equation}

In terms of each optical element's reflectance  and  transmittance, the response factor reads

\begin{equation}
\label{eq:LIGO_response}
     \hspace*{-1cm}\Gamma =  \frac{-t_1r_2t_2r_3\e^{ikd_0}\bigg[\e^{2ikd_1}\bigg(1-\frac{t_3^2\e^{2ikd_2}}{1-r_3^2\e^{2ikd_2}}\bigg) + \e^{2ikd_3}\bigg(1-\frac{t_3^2\e^{2ikd_4}}{1-r_3^2\e^{2ikd_4}}\bigg)\bigg]}{1 - r_1r_3\e^{2ikd_0}\bigg[ r_2^2\e^{2ikd_1}\bigg( 1-\frac{t_3^2\e^{2ikd_2}}{1-r_3^2\e^{2ikd_2}}\bigg)-t_2^2\e^{2ikd_3}\bigg(1 - \frac{t_3^2\e^{2ikd_4}}{1-r_3^2\e^{2ikd_4}} \bigg) \bigg]} .
\end{equation}

\section{Intermediate fields}
The calculation of the resultant electric field as a superposition of wave fronts undergoing different optical paths can be extended to positions other than the output. By treating an arbitrary vertex as an output and simplifying the graph so that only the input vertex and this arbitrary one remain, the response factor for the electric field in an arbitrary position can be calculated.

This is useful when examining, for example, the difference between the power stored in each side of a cavity containing a membrane in the middle \cite{Thompson2008}, the trapping power acting on particles trapped by the stationary wave of an optical cavity \cite{Kiesel2013} and the power stored in a cavity-enhanced Michelson interferometer \cite{Martynov2017}. The latter interferometer will now be used as an example.

First, consider the response factor in (\ref{eq:LIGO_response}). It is useful to add some constraints based on the operation of this kind of interferometer \cite{paper_LIGO_readout}. First of all, both Fabry-P\'erot cavities must be kept in resonance, which leads to
\begin{eqnarray}
\label{eq:resonance_condition}
    d_2 =  n\frac{\lambda}{2} \;,\; &&\;\; 
    d_4 = m\frac{\lambda}{2} \, ,
\end{eqnarray}
with $m,n\in \mathbb{N^*}$. In addition to that, the output of the Michelson interferometer must be kept dark, that is, $\vert \Gamma \vert^2$ must be minimum. Provided that the BS is balanced ($r_2=t_2=1/\sqrt{2}$), the minimum absolute value of (\ref{eq:LIGO_response}) is reached for

\begin{equation}
\label{eq:dark_michelson}
    d_1 = d_3+\left(p+\frac{1}{2}\right)\frac{\lambda}{2}
\end{equation}
with $p\in\mathbb{Z}$.

\begin{figure}[t]
    \centering
    \includegraphics[scale = 0.8]{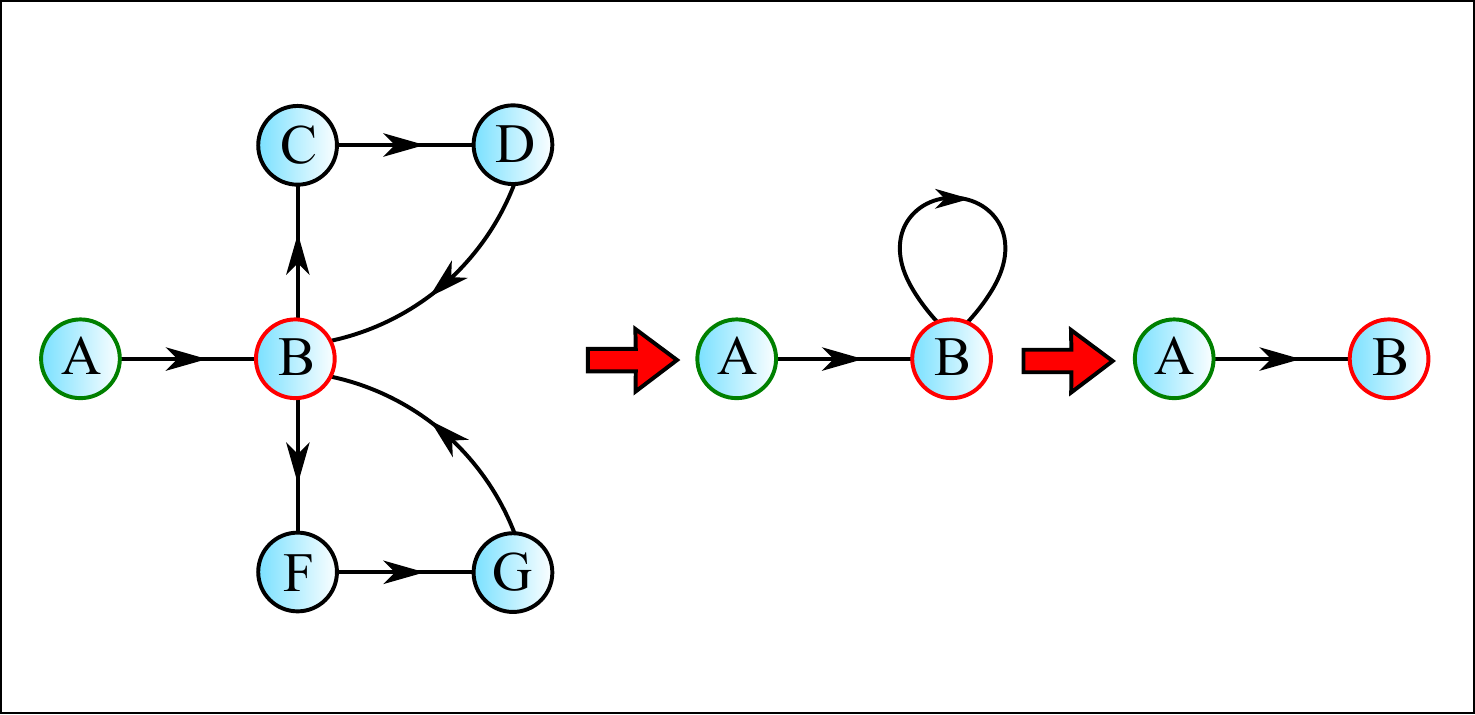}
    \caption{Simplification of a graph corresponding to the cavity-enhanced Michelson interferometer when the resultant electric field in the intermediate state B is of interest. The rules are applied until only vertices A and B are left. The weight of the edge connecting A to B is equal to the response factor $\Gamma_{AB}$ that relates the input electric field to the electric field at B.}
    \label{fig:LIGO_Intermediary}
\end{figure}

Finally, the PRM must be in a position that causes the power stored in the arms of the interferometer to build up. This can be achieved by maximizing the intensity of the wave travelling to the right after the PRM, that is the resultant wave-front in state B. This resultant electric field can be calculated by finding the response factor $\Gamma_{\rm{AB}}$, given by the sum of the weights of all walks in the graph of figure \ref{fig:LIGO_diagram} that start at A and end at B. Figure \ref{fig:LIGO_Intermediary} shows the successive application of simplification rules leading to a graph containing only vertices A and B, which corresponds to the input state and the state in which the resultant electric field is to be calculated. 

First, erase the vertex I, since there is no walk that passes by this vertex and ends at B. Then apply a loop contraction between C and D [F and G] and sum the resultant parallel edges, yielding an edge from C to D [F to G] with weight $\Phi_{\rm{CD},2} = \Phi_{\rm{CD}}+\Phi_{\rm{CE}}\Phi_{\rm{ED}}/(1-\Phi_{\rm{EE}})$ [$\Phi_{\rm{FG},2} = \Phi_{\rm{FG}}+\Phi_{\rm{FH}}\Phi_{\rm{HG}}/(1-\Phi_{\rm{HH}})$]. Next, contract all consecutive edges and sum the two loops in $B$, resulting in a single loop in $B$ with weight $\Phi_{\rm{BB}}=\Phi_{\rm{BC}}\Phi_{\rm{CD},2}\Phi_{\rm{DB}}+\Phi_{\rm{BF}}\Phi_{\rm{FG},2}\Phi_{\rm{GB}}$.

The presence of a loop in  vertex B is a direct consequence of treating a vertex that is not a sink vertex as an output. This loop can be contracted following a calculation similar to the one used for the Fabry-Per\'ot interferometer, yielding

\begin{equation}
    \Gamma_{\rm{AB}} = \frac{\Phi_{\rm{AB}}}{1-\Phi_{\rm{BB}}},
\end{equation}
or, substituting the weights together with conditions  (\ref{eq:resonance_condition}) and (\ref{eq:dark_michelson}),
\begin{equation}
\label{eq:LIGO_intermediate}
     \hspace*{-1cm}\Gamma_{\rm{AB}} =  \frac{-t_1r_2t_2r_3\e^{ikd_0}}{1 - r_1r_3\e^{2ikd_0}\bigg[ r_2^2\e^{2ikd_1}\bigg( 1-\frac{t_3^2\e^{2ikd_2}}{1-r_3^2\e^{2ikd_2}}\bigg)-t_2^2\e^{2ikd_3}\bigg(1 - \frac{t_3^2\e^{2ikd_4}}{1-r_3^2\e^{2ikd_4}} \bigg) \bigg]} .
\end{equation}

Maximizing $\vert \Gamma_{\rm{AB}} \vert$, with $d_1$, $d_2$, $d_3$ and $d_4$ satisfying the conditions in (\ref{eq:resonance_condition}) and (\ref{eq:dark_michelson}) yields
\begin{eqnarray}
\label{eq:PRM_1}
    d_0+d_1 = q\frac{\lambda}{2}\nonumber\\
    d_0+d_3 = s\frac{\lambda}{2},
\end{eqnarray}

\noindent with $q,s\in\mathbb{N^*}$. The condition in (\ref{eq:PRM_1}) is exactly the one found by imposing that the PRM together with the input mirrors of the Fabry-P\'erot cavities form two resonant cavities, as expected \cite{paper_LIGO}.

Note that, in this example, only the amplitude of the resultant wave travelling to the right in figure \ref{fig:LIGO_schematics} has been calculated. If the intensity of the standing wave was to be found, the amplitude of the wave travelling to the left at that same arm of the interferometer would have to be calculated. This could be easily done by defining an extra state at the same position as B, but with opposite direction.

\section{Multiple inputs and outputs} \label{Multiple_IOs}
\begin{figure}[t]
\centering
\subcaptionbox{\ \label{fig:Mach_Zehnder_schematics}}
{\includegraphics[scale = 0.8]{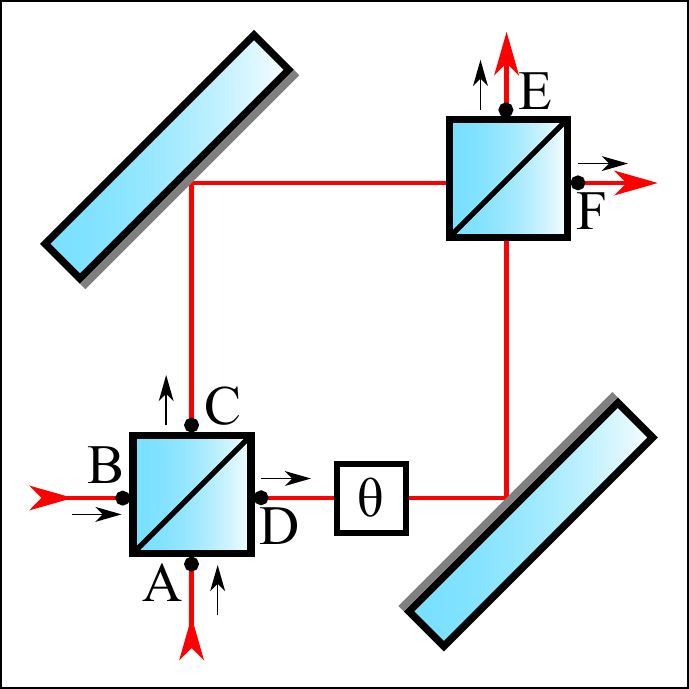}}
\subcaptionbox{\ \label{fig:Mach_Zehnder_diagram}}
{\includegraphics[scale = 0.8]{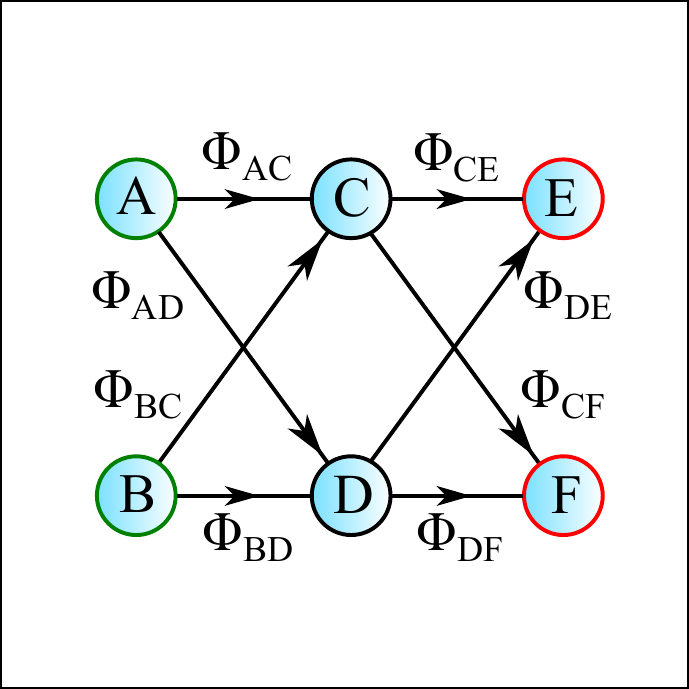}}
\caption{(a) Optical schematics corresponding to a Mach-Zehnder interferometer. Differences between the two paths inside the interferometer are summarized by a phase $\theta$. (b) Graph corresponding to the Mach-Zehnder interferometer, with two input states, indicated by the green circles, and two output states, indicated by the red circles.}\label{fig:Mach_Zehnder}
\end{figure} 

Up to this point solely interferometers with one input and one output have been studied, but the presented graph-based method can be easily generalized for homodyne systems containing an arbitrary number of inputs and outputs.

As discussed in the end of section \ref{sec:rules}, the simplification rules are local, and, therefore, must remain valid regardless of the number of I/O vertices. A difference appears at each output, where one must sum over the fields coming from each input port of the interferometer. In terms of the response factors, the electric field in a given output $\rm{O}_M$ is
\begin{equation}
    \vec{E}_{O_M} = \sum_n \Gamma_{nM}\vec{E}_{I_n}
\end{equation}
where $n$ must run over all input ports.

The following example serves to illustrate how to find the transmitted field through an interferometer with multiple inputs/outputs. The Mach–Zehnder interferometer displayed in figure \ref{fig:Mach_Zehnder_schematics} has two BS's, with reflectance and transmittance $r, t$ as well as two perfect mirrors. The phase shift between each arm of the interferometer is taken to be $\theta$, due to a phase shifter.

States A through F are defined, giving rise to the graph in figure \ref{fig:Mach_Zehnder_diagram}, with the following weights
\begin{eqnarray}
\Phi_{\rm{AC}} = it &&\;\; \Phi_{\rm{AD}} = r \nonumber\\
\Phi_{\rm{BC}} = r &&\;\; \Phi_{\rm{BD}} = it \nonumber\\
\Phi_{\rm{CE}} = r && \;\;\Phi_{\rm{CF}} = it \nonumber\\
\Phi_{\rm{DE}} = it\e^{i\theta} &&\;\; \Phi_{\rm{DF}} = r\e^{i\theta} \label{eq:weights_mach_zehnder}
\end{eqnarray}

\begin{figure}[t]
    \centering
    \includegraphics[scale = 0.8]{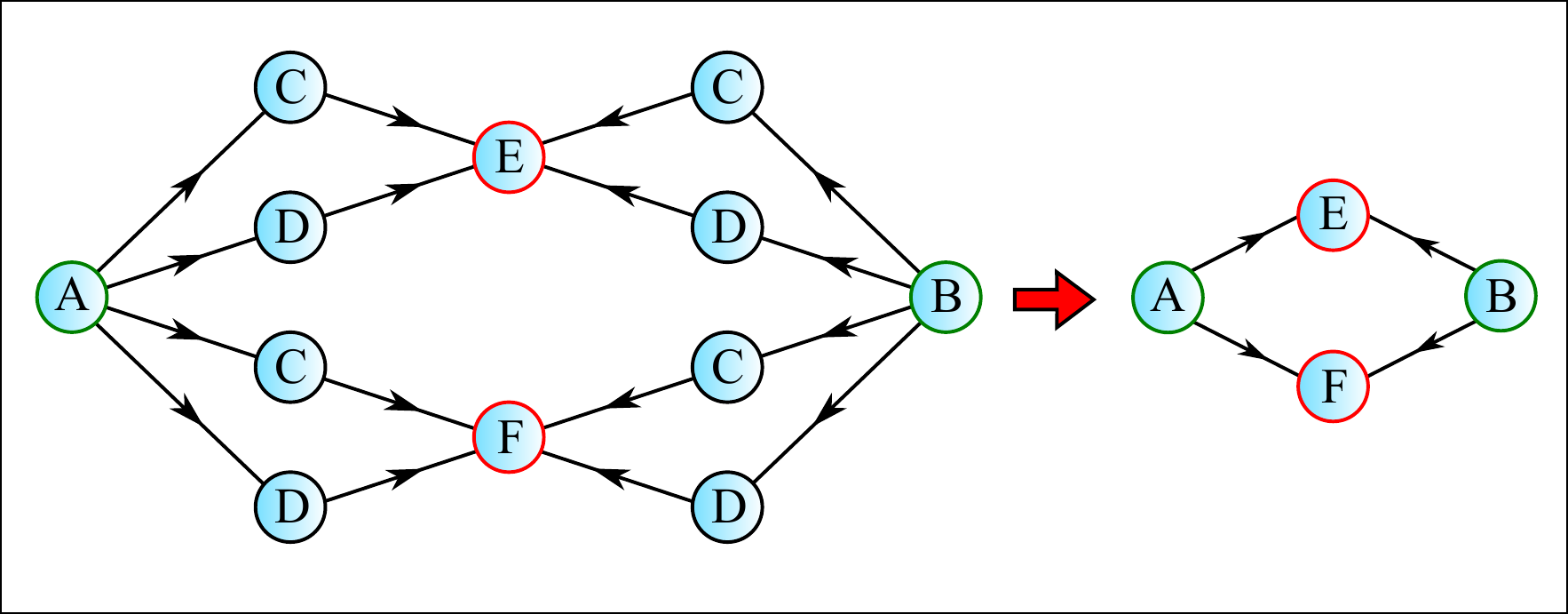}
    \caption{Simplification of the graph corresponding to a Mach-Zehnder interferometer. The graph is simplified until all that remains are the input and output vertices.}\label{fig:Mach_Zehnder_simplification}
\end{figure}

To find the transmitted field through the interferometer, first one needs to detach the edges on vertices C and D and then replace the resulting consecutive and parallel edges. This sequence of operation is shown in figure \ref{fig:Mach_Zehnder_simplification}.

The output electric field at the output state E [F] will be the sum of the output electric field in E [F] coming from state A and the output electric field in E [F] coming from state B
\begin{eqnarray}
    \vec{E}_{\rm{out, E}} = \big(\Phi_{\rm{AC}}\Phi_{\rm{CE}} + \Phi_{\rm{AD}}\Phi_{\rm{DE}}\big)\vec{E}_{\rm{in,A}} + \big(\Phi_{\rm{BC}}\Phi_{\rm{CE}} + \Phi_{\rm{BD}}\Phi_{\rm{DE}}\big)\vec{E}_{\rm{in,B}} \label{mach_zehnder_output_E} \, \nonumber\\
    \vec{E}_{\rm{out, F}} = \big(\Phi_{\rm{AC}}\Phi_{\rm{CF}} + \Phi_{\rm{AD}}\Phi_{\rm{DF}}\big)\vec{E}_{\rm{in,A}} + \big(\Phi_{\rm{BC}}\Phi_{\rm{CF}} + \Phi_{\rm{BD}}\Phi_{\rm{DF}}\big)\vec{E}_{\rm{in,B}} \label{mach_zehnder_output_F} \, .
\end{eqnarray}

Plugging the edge weights  from (\ref{eq:weights_mach_zehnder}) into  (\ref{mach_zehnder_output_E})  the standard result is obtained
\begin{eqnarray}
\vec{E}_{\rm{out, E}} = irt\bigg(1+\e^{i\theta}\bigg)\vec{E}_{\rm{in,A}} + \bigg(r^2-t^2\e^{i\theta}\bigg)\vec{E}_{\rm{in,B}} \, , \nonumber \\
\vec{E}_{\rm{out, F}}=\bigg(r^2\e^{i\theta}-t^2\bigg)\vec{E}_{\rm{in,A}} + irt\bigg(1+\e^{i\theta}\bigg)\vec{E}_{\rm{in,B}} \, 
.
\end{eqnarray}

By following the same procedure one can calculate the electric field at any output in the previous examples. One could, for example, define a state D in figure \ref{fig:Fabry_Schematics} at the same position as state A but with opposite direction. Simplification of the associated graph would lead to an edge having weight $\Phi_{\rm{AC}}$, equal to the response factor for the cavity's transmission, and an edge with weight $\Phi_{\rm{AD}}$, equal to the response factor for the cavity's reflection.

\section{Quantum interferometry}
The present graph-based method can also be used to describe interferometers whose inputs are not classical electric fields, but quantum states of light \cite{Scully1997}.

So far the effect of an interferometer upon input electric fields transforming into output electric fields has been described. With second quantization, the same principle can be applied to unravel how the input modes' creation/annihilation operators evolve to the output modes' creation/annihilation operators.

In order to do so, note that the response factors carry information about how much of the light that enters an interferometer trough one port leaves it trough another, as well as the phase gained by doing so. This can be described mathematically by a transformation of the form

\begin{equation}
    \hat{a}^\dagger \rightarrow \Gamma_{\rm{AB}}\,\hat{b}^\dagger + \Gamma_{\rm{AC}}\,\hat{c}^\dagger + \cdots \, , \label{quantum_transformation}
\end{equation}
\noindent where $\hat{a}^\dagger$ is the creation operator associated with the input state A; $\hat{b}^\dagger, \hat{c}^\dagger,\ldots$ are the creation operators associated with the output states B, C, ... and $\Gamma_{\rm{AB}}$, $\Gamma_{{\rm{A}}C}$, ... are the response factors between state A and states B, C, and so on.

When treating quantum states of light, all possible optical paths must be considered, otherwise, the evolution of the input creation operators will not be unitary. This is in contrast with the classical case, in which paths that are not of interest might be ignored.

As a final remark, any multi-mode quantum state in the Fock basis can be written as a function of each mode's creation and annihilation operators acting on the vacuum state
\begin{equation}
 \vert \psi \rangle = \hat{\Psi}(\hat{a}, \hat{b}, \hat{c}, ..., \hat{a}^{\dagger}, \hat{b}^{\dagger}, \hat{c}^{\dagger}, ... ) \vert 0 \rangle \label{eq:arb_state}
\end{equation}

\noindent and so, the effect of the interferometer upon an arbitrary input quantum state can be obtained by transforming the input modes according to the interferometer.

\subsection{Hong-Ou-Mandel effect}
\begin{figure}[t]
\centering
\subcaptionbox{\ \label{fig:BS_Schematics}}
{\includegraphics[scale = 0.8]{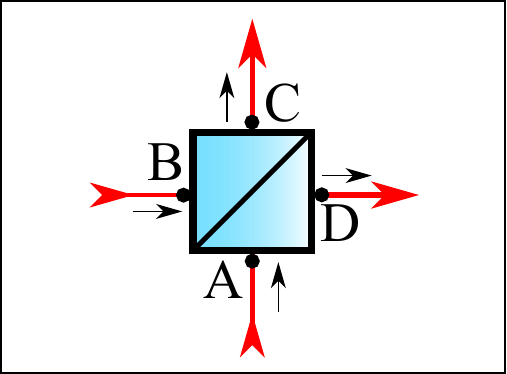}}
\subcaptionbox{\ \label{fig:BS_Diagram}}
{\includegraphics[scale = 0.8]{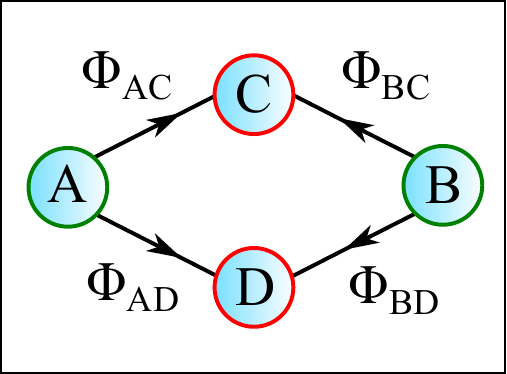}}
\caption{(a) Optical schematics corresponding to a beam splitter. (b) Graph corresponding to the beam splitter. The graph is already in its simplest form, which means that the response factor between a given input and output is simply the weight of the edge connecting the corresponding vertices.}\label{fig:BS}
\end{figure}

Consider the example of a balanced BS ($r=t=1/\sqrt{2}$), shown in figure \ref{fig:BS_Schematics}, with states A, B standing for the inputs and C, D for the outputs. The associated graph, shown in figure \ref{fig:BS_Diagram} is already on its simplest form, but it serves to illustrate how one can deal with the effects of quantum statistics. Let 
 $a^{\dagger}$, $b^{\dagger}$, $c^{\dagger}$ and $d^{\dagger}$ be the operators that create a single photon in the states A, B, C and D, respectively. The input operators transform into the output operators according to
\begin{eqnarray}
    a^{\dagger} &\rightarrow \Gamma_{\rm{AC}}\, c^{\dagger} + \Gamma_{\rm{AD}}\, d^{\dagger}\, ,\nonumber \\ 
     b^{\dagger}&\rightarrow \Gamma_{\rm{BD}}\, d^{\dagger} + \Gamma_{\rm{BC}}\, c^{\dagger} \, ,
\end{eqnarray}
where, in this simple example, $ \Gamma_{\rm{AC}} = \Gamma_{\rm{BD}} = i/ \sqrt{2} $ and  $ \Gamma_{\rm{AD}} = \Gamma_{\rm{BC}} = 1 / \sqrt{2} $. If the input are two indistinguishable photons, one in each port of the BS, the famous Hong-Ou-Mandel effect, characteristic of bosonic statistics, arises \cite{HOM1987}
\begin{equation}
    a^{\dagger} b^{\dagger} \rightarrow \Gamma_{\rm{AC}}\Gamma_{\rm{BC}}\, {c^{\dagger}}^{2} + \Gamma_{\rm{AD}}\Gamma_{\rm{BD}}\, {d^{\dagger}}^{2} = 
    \frac{i}{2}({c^{\dagger}}^{2}+{d^{\dagger}}^{2}) \, .
\end{equation}

\subsection{Quantum states in a Mach-Zehnder interferometer}
\indent The Mach-Zehnder interferometer, whose schematics and associated graph are represented in figure \ref{fig:Mach_Zehnder}, is now revisited. From section \ref{Multiple_IOs}, it is known that the response factors are

\begin{eqnarray}
\Gamma_{\rm{AE}} = irt\bigg(1+\e^{i\theta}\bigg) \;\; && \;\; \Gamma_{\rm{BE}} =  \bigg(r^2-t^2\e^{i\theta}\bigg)\, ,\nonumber \\ \Gamma_{\rm{AF}} = \bigg(r^2\e^{i\theta}-t^2\bigg)  \;\; && \;\; \Gamma_{\rm{BF}} = irt\bigg(1+\e^{i\theta}\bigg)\, .
\end{eqnarray}

Thus, the evolution of the input modes' creation operators are easily found

\begin{eqnarray} 
\hat{a}^\dagger \rightarrow irt\bigg(1+\e^{i\theta}\bigg)\hat{e}^\dagger + \bigg(r^2\e^{i\theta}-t^2\bigg)\hat{f}^\dagger\, , \label{QMZ_transformation_a}\nonumber\\
\hat{b}^\dagger \rightarrow \bigg(r^2-t^2\e^{i\theta}\bigg)\hat{e}^\dagger + irt\bigg(1+\e^{i\theta}\bigg)\hat{f}^\dagger\, , \label{QMZ_transformation_b}
\end{eqnarray}

\noindent which is the well-known result for a quantum Mach-Zehnder interferometer \cite{Kim03}.

\section{Conclusion}
In summary, we have shown how to associate a directed weighted graph to an interferometric setup. Using the so-called simplification rules, it is possible to transform the directed graph, as to simplify it as much as possible, and get the response factors for the interferometer's outputs, as well as for any point in the setup. From these response factors, the electric field, as well as quantum states of light in the Fock basis, within an interferometer can be obtained.

To illustrate the technique, several examples were analysed, such as the Michelson and Fabry-P\'erot interferometers, the optomechanical setup of multiple membranes inside an optical cavity, the cavity-enhanced Michelson interferometer and the Mach-Zehnder interferometer. The graphical approach provides a clear physical picture in contrast to the standard transfer-matrix approach. It also translates the physical problem of calculating the electrical field in an interferometer to a combinatorial problem. It will be interesting to explore, in future works, the connections between the calculus of amplitudes and response factors to the mathematical theory of directed graphs, as well as the extension of the method to heterodyne optical systems.

\ack
The authors acknowledge Leandro Lyra and Peter Gibson for useful discussions. This study was financed in part by the Coordenac\~ao de
Aperfei\c{c}oamento de Pessoal de N\'ivel Superior - Brasil
(CAPES) - Finance Code 001 and by  CNPq (Conselho Nacional de Desenvolvimento Cient\'ifico e Tecnol\'ogico).

\newcommand{\newblock}{}

\bibliography{main}

\providecommand{\newblock}{}
\begin{thebibliography}{10}
\expandafter\ifx\csname url\endcsname\relax
  \def\url#1{{\tt #1}}\fi
\expandafter\ifx\csname urlprefix\endcsname\relax\def\urlprefix{URL }\fi
\providecommand{\eprint}[2][]{\url{#2}}

\bibitem{Curtis1998}
Curtis E, Ingerman D and Morrow J 1998 {\em Linear Algebra and its
  Applications\/} {\bf 283} 115 -- 150 ISSN 0024-3795

\bibitem{Gutman1986}
Gutman I and Polansky O 1986 {\em Mathematical concepts in organic chemistry\/}
  (Springer-Verlag)

\bibitem{Rouvray1996}
Rouvray D~H 1996 {\em Combinatorics in Chemistry\/} (The MIT Press) p
  1955–1981

\bibitem{Feynman1965}
Feynman R~P and Hibbs A~R 1965 {\em Quantum Mechanics and Path Integrals\/}
  (McGraw-Hill College)

\bibitem{Helvang2013}
Elvang H and Huang Y~T 2013 Scattering amplitudes (\textit{Preprint}
  \eprint{1308.1697})

\bibitem{kauffman1991}
Kauffman L~H 1991 {\em Knots and Physics\/} (World Scientific)

\bibitem{Kardar2007}
Kardar M 2007 {\em Statistical Physics of Fields\/} (Cambridge University
  Press)

\bibitem{Albert2002}
Albert R and Barabási A~L 2002 {\em Reviews of Modern Physics\/} {\bf 74}
  47–97

\bibitem{Nielsen2006}
Nielsen M~A 2006 {\em Reports on Mathematical Physics\/} {\bf 57} 147–161

\bibitem{Rossi2013}
Rossi M, Huber M, Bru{\ss} D and Macchiavello C 2013 {\em New Journal of
  Physics\/} {\bf 15} 113022

\bibitem{Gibson2018}
Gibson P~C 2018 {\em Journal of Computational Physics\/} {\bf 372} 524--545

\bibitem{Gibson2019}
Gibson P~C 2019 {\em Journal of Approximation Theory\/} {\bf 244} 37--56

\bibitem{Krenn_2017}
Krenn M, Gu X and Zeilinger A 2017 {\em Phys. Rev. Lett.\/} {\bf 119}(24)
  240403

\bibitem{Gu_2019}
Gu X, Erhard M, Zeilinger A and Krenn M 2019 {\em Proceedings of the National
  Academy of Sciences\/} {\bf 116} 4147--4155 ISSN 0027-8424

\bibitem{Ataman2015ThreeExperiments}
Ataman S 2015 {\em The European Physical Journal D\/} {\bf 69} 44 ISSN
  1434-6079 \urlprefix\url{https://doi.org/10.1140/epjd/e2014-50693-1}

\bibitem{Ataman2015Fabry}
Ataman S 2015 {\em The European Physical Journal D\/} {\bf 69} 187 ISSN
  1434-6079 \urlprefix\url{https://doi.org/10.1140/epjd/e2015-60211-8}

\bibitem{Ataman_2018}
Ataman S 2018 {\em Journal of Physics Communications\/} {\bf 2} 035032

\bibitem{Ataman2014First}
Ataman S 2014 {\em The European Physical Journal D\/} {\bf 68} 288 ISSN
  1434-6079 \urlprefix\url{https://doi.org/10.1140/epjd/e2014-50448-0}

\bibitem{paper_LIGO}
Abbott B~P {\em et~al.\/} 2009 {\em Reports on Progress in Physics\/} {\bf 72}
  076901

\bibitem{Aspelmeyer2014}
Aspelmeyer M, Kippenberg T~J and Marquardt F 2014 {\em Reviews of Modern
  Physics\/} {\bf 86} 1391–1452

\bibitem{Newsom2019}
Newsom D~C, Luna F, Fedoseev V, Löffler W and Bouwmeester D 2019
  (\textit{Preprint} \eprint{arXiv:1909.11384v1})

\bibitem{Bhattacharya2008}
Bhattacharya M and Meystre P 2008 {\em Phys. Rev. A\/} {\bf 78}(4) 041801

\bibitem{Saleh2007}
B~E A~Saleh M~C~T 2007 {\em Fundamentals of Photonics\/} (Wiley)

\bibitem{Osorio2012}
Osorio C~I, Bruno N, Sangouard N, Zbinden H, Gisin N and Thew R~T 2012 {\em
  Phys. Rev. A\/} {\bf 86}(2) 023815

\bibitem{Sangouard2011}
Sangouard N, Simon C, de~Riedmatten H and Gisin N 2011 {\em Rev. Mod. Phys.\/}
  {\bf 83}(1) 33--80

\bibitem{Tillman2015}
Tillmann M, Tan S~H, Stoeckl S~E, Sanders B~C, de~Guise H, Heilmann R, Nolte S,
  Szameit A and Walther P 2015 {\em Phys. Rev. X\/} {\bf 5}(4) 041015

\bibitem{Krenn2016}
Krenn M, Malik M, Fickler R, Lapkiewicz R and Zeilinger A 2016 {\em Phys. Rev.
  Lett.\/} {\bf 116}(9) 090405

\bibitem{Jayich2008}
Jayich A~M, Sankey J~C, Zwickl B~M, Yang C, Thompson J~D, Girvin S~M, Clerk
  A~A, Marquardt F and Harris J~G~E 2008 {\em New Journal of Physics\/} {\bf
  10} 095008

\bibitem{paper_2_membranes}
Piergentili P, Catalini L, Bawaj M, Zippilli S, Malossi N, Natali R, Vitali D
  and Giuseppe G~D 2018 {\em New Journal of Physics\/} {\bf 20} 083024

\bibitem{Ismail2016}
Ismail N, Kores C~C, Geskus D and Pollnau M 2016 {\em Optics Express\/} {\bf
  24} 16366

\bibitem{Webb1998a}
Webb M~S, Moulton P~F, Kasinski J~J, Burnham R~L, Loiacono G and Stolzenberger
  R 1998 {\em Optics Letters\/} {\bf 23} 1161

\bibitem{Sato2000}
Sato S, Ohashi M, Fujimoto M, Fukushima M, Waseda K, Miyoki S, Mavalvala N and
  Yamamoto H 2000 {\em Applied Optics\/} {\bf 39} 4616

\bibitem{Wei2019}
Wei X, Sheng J, Wu Y, Liu W and Wu H 2019 {\em Applied Physics Letters\/} {\bf
  115} 251105

\bibitem{Magrini2018}
Magrini L, Norte R~A, Riedinger R, Marinkovi{\'{c}} I, Grass D, Deli{\'{c}} U,
  Gröblacher S, Hong S and Aspelmeyer M 2018 {\em Optica\/} {\bf 5} 1597

\bibitem{Vitali2002}
Vitali D, Mancini S, Ribichini L and Tombesi P 2002 {\em Phys. Rev. A\/} {\bf
  65}(6) 063803

\bibitem{Thompson2008}
Thompson J~D, Zwickl B~M, Jayich A~M, Marquardt F, Girvin S~M and Harris J~G~E
  2008 {\em Nature\/} {\bf 452} 72--75

\bibitem{Li2016}
Li J, Xuereb A, Malossi N and Vitali D 2016 {\em Journal of Optics\/} {\bf 18}
  084001

\bibitem{Xu2013}
Xu X~W, Zhao Y~J and Liu Y~X 2013 {\em Phys. Rev. A\/} {\bf 88}(2) 022325

\bibitem{Muler2003}
Müler G, Delker T, Tanner D~B and Reitze D 2003 {\em Applied Optics\/} {\bf
  42} 1257

\bibitem{Thuring2005}
Th\"uring A, L\"uck H and Danzmann K 2005 {\em Phys. Rev. E\/} {\bf 72}(6)
  066615

\bibitem{Abramovici1992}
Abramovici A, Althouse W~E, Drever R~W~P, Gursel Y, Kawamura S, Raab F~J,
  Shoemaker D, Sievers L, Spero R~E, Thorne K~S, Vogt R~E, Weiss R, Whitcomb
  S~E and Zucker M~E 1992 {\em Science\/} {\bf 256} 325--333

\bibitem{Kiesel2013}
Kiesel N, Blaser F, Delic U, Grass D, Kaltenbaek R and Aspelmeyer M 2013 {\em
  Proceedings of the National Academy of Sciences\/} {\bf 110} 14180--14185

\bibitem{Martynov2017}
Martynov {\em et~al.\/} (LSC Instrument Authors) 2017 {\em Phys. Rev. A\/} {\bf
  95}(4) 043831

\bibitem{paper_LIGO_readout}
Fritschel P, Bork R, Gonz\'{a}lez G, Mavalvala N, Ouimette D, Rong H, Sigg D
  and Zucker M 2001 {\em Appl. Opt.\/} {\bf 40} 4988--4998

\bibitem{Scully1997}
Scully M~O and Zubairy M~S 1997 {\em Quantum Optics\/} (Cambridge University
  Press)

\bibitem{HOM1987}
Hong C~K, Ou Z~Y and Mandel L 1987 {\em Phys. Rev. Lett.\/} {\bf 59}(18)
  2044--2046

\bibitem{Kim03}
Kim T~S, Kim H~O, Ko J~H and Park G~D 2003 {\em J. Opt. Soc. Korea\/} {\bf 7}
  113--118

\end{thebibliography}

\end{document}